\documentclass[apj]{emulateapj} 
\usepackage{lscape} 
\usepackage{apjfonts}
\usepackage{latexsym}
\usepackage{graphicx,natbib}
\usepackage{hyperref}
\usepackage{amsmath} 
\usepackage{subfigure}

\shorttitle{}
\shortauthors{A. Monachesi et al.}
\begin{document}

\title{Testing galaxy formation models with the GHOSTS survey: The
  color profile of M81's stellar halo}

\author{Antonela Monachesi\altaffilmark{1}}\email{antonela@umich.edu}
\author{Eric F. Bell\altaffilmark{1}}\author{David
  J. Radburn-Smith\altaffilmark{2}}\author{Marija
  Vlaji\'c\altaffilmark{3}} \author{Roelof S. de
  Jong\altaffilmark{3}}\author{Jeremy Bailin\altaffilmark{1}}
\author{Julianne J. Dalcanton\altaffilmark{2}}\author{Benne W.
  Holwerda\altaffilmark{4}}\author{David Streich\altaffilmark{3}}

  \altaffiltext{1}{Department of Astronomy, University of Michigan,
    830 Dennison Bldg., 500 Church St., Ann Arbor, MI 48109, USA}
  \altaffiltext{2}{Department of Astronomy, University of Washington,
    Seattle, WA 98195,USA} \altaffiltext{3}{Leibniz-Institut f\"ur
    Astrophysik Potsdam, D-14482 Potsdam, Germany}
  \altaffiltext{4}{European Space Agency Research Fellow (ESTEC),
    Keplerlaan 1, 2200 AG Noordwijk, The Netherlands}
  \footnotetext[5]{Based on observations made with the NASA/ESA Hubble
    Space Telescope, obtained at the Space Telescope Science
    Institute, which is operated by the Association of Universities
    for Research in Astronomy, Inc., under NASA contract NAS 5-26555.}

\begin{abstract}
 We study the properties of the stellar populations in M81's outermost
 part, which hereafter we will term the stellar halo, using {\it
   Hubble Space Telescope} (HST) Advanced Camera for Surveys
 observations of 19 fields from the GHOSTS survey. The observed fields
 probe the stellar halo out to a projected distance of $\sim$ 50 kpc
 from the galactic center. Each field was observed in both $F606W$ and
 $F814W$ filters. The 50\% completeness levels of the color magnitude
 diagrams (CMDs) are typically at 2 mag below the tip of the red giant
 branch (TRGB). Fields at distances closer than 15 kpc show evidence
 of disk-dominated populations whereas fields at larger distances are
 mostly populated by halo stars. The RGB of the M81's halo CMDs is
 well matched with isochrones of $\sim$ 10 Gyr and metallicities
 $[\mathrm{Fe/H}] \sim -1.2$ dex, suggesting that the dominant stellar
 population of M81's halo has a similar age and metallicity. The halo
 of M81 is characterized by a color distribution of width $\sim 0.4$
 mag and an approximately constant median value of ($F606W - F814W$)
 $\sim 1$ mag measured using stars within the magnitude range $23.7
 \lesssim F814W \lesssim 25.5$. When considering only fields located
 at galactocentric radius $R > 15$ kpc, we detect no color gradient in
 the stellar halo of M81. We place a limit of $0.03\pm0.11$ mag
 difference between the median color of RGB M81 halo stars at $\sim$15
 and at 50 kpc, corresponding to a metallicity difference of
 $0.08\pm0.35$ dex over that radial range for an assumed constant age
 of 10 Gyr. We compare these results with model predictions for the
 colors of stellar halos formed purely via accretion of satellite
 galaxies. When we analyze the cosmologically motivated models in the
 same way as the HST data, we find that they predict no color gradient
 for the stellar halos, in good agreement with the observations.
\end{abstract}


\maketitle

\section{Introduction}\label{sec:intro}

According to the currently favored cosmological model, $\Lambda$CDM,
galaxies form hierarchically through the accretion of smaller objects
that, due to gravity, merge together to form the larger systems we see
today \citep[e.g.][]{White_Rees78}. Predictions of this theory, such
as the existence of coherent streams, shells, satellite galaxies, and
other substructure in the halos of galaxies \citep{Johnston96,
  Helmi_White99, Helmi_dezeeuw00, Bullock01, BJ05, DeLucia_helmi08,
  Gomez10b, Cooper10} have been supported by observations not only in
our own Galaxy \citep{Ibata95, Helmi99, Majewski03, Yanny03, Ibata03,
  Belokurov06, Juric08, Bell08, Grillmair09}, but also in M31
\citep{Ibata07,McConnachie09}, and in more distant
galaxies~\citep[e.g.][]{Martinezdelgado09, Martinezdelgado10,
  Mouhcine10, Bailin11}.

Another important prediction provided by theoretical models is the
existence of stellar population variations within stellar halos
\citep{Robertson05,Font06, Font08, Cooper10, Tumlinson10}. Metallicity
variations are expected due to the mass-metallicity relation assumed
for the accreted satellite galaxies in the models, such that more
massive satellites have higher metallicities. Age variations are also
expected due to the differences in the satellite accretion times.

The halo of our own Galaxy has been extensively studied and variations
of its stellar populations have been detected
\citep[e.g.,][]{Ivezic08}. \citet{Bell10} have used the ratio of blue
horizontal branch stars (BHB) to main sequence turn off stars (MSTO)
from the Sloan Digital Sky Survey \citep[SDSS][]{York00} to show that
large stellar population variations can be detected out to
heliocentric distances of $\sim 30$ kpc. Moreover, they found that
these variations trace different previously identified structures. In
M31, variations in the halo's stellar populations have also been
detected using deep Hubble Space Telescope (HST) observations
\citep{Brown06, Richardson08, Richardson09, Sarajedini12}. These
variations were associated with substructures detected from ground
based observations \citep{Ibata07, McConnachie09}.

Even though stellar population variations are present throughout the
stellar halos of both the Milky Way and M31, the existence of
metallicity gradients is still a matter of debate. For the Milky Way,
recent studies have analyzed different stellar samples from the SDSS;
\citep{Carollo07, Carollo10} detected variations on the mean
metallicity with distance from the Galactic plane, with stars at $R<
15$ kpc more metal-rich ($[\textrm{Fe/H}] \sim -1.6$) than those at
larger radii ($R > 15$--20 kpc) with an average metallicity of
$[\textrm{Fe/H}] \sim -2.2$ dex. They considered a sample of halo
stars located in the Solar Neighborhood, but reaching the outer
regions of the halo in their orbits. They suggested the existence of
two distinct components of the Halo, i.e. inner and outer
halo. \citet{Ivezic08} used a large sample of stars that reach to 8
kpc from the Sun and found a mean metallicity of $[\textrm{Fe/H}] \sim
-1.46$ dex with a nearly flat radial metallicity gradient. They assume
that their sample, however, is likely to be dominated by inner halo
stars. Using photometric data from SDSS/SEGUE survey, \citet{deJong10}
constructed a color magnitude diagram (CMD) of Milky Way thick disk
and halo stars that are at $|Z| > 1$ kpc from the Galactic plane. They
adopted a single age of $\sim 14$ Gyr and three metallicity bins of
$[\textrm{Fe/H}] = -0.7$, $-1.3$, and $-2.2$ to fit the turnoff stars
of their CMD, presuming that those three components are sufficient to
fully describe both the thick disk and halo stars of their
sample. They found a mean halo metallicity of $[\textrm{Fe/H}] = -1.6$
at distances smaller than 10 kpc and a mean metallicity of
$[\textrm{Fe/H}] = -2.2$ for the stellar halo at distances larger than
15 kpc, consistent with the results of \citet{Carollo07}. A more
recent study, however, utilizes an 'in situ' sample of K giants from
SDSS/SEGUE spectroscopic observations and show a nearly flat Galactic
halo metallicity distribution ($[\textrm{Fe/H}] \sim -1.4$ dex) within
$\sim 20$ to $\sim60$ kpc \citep[Ma et al. in prep.]{Ma12}.  We
recognize, nevertheless, that all the mentioned studies may have
important biases introduced by, e.g., the magnitude or color limit
considered, which affect the determination of a metallicity--distance
relation and need to be carefully taken into account \citep[see
  e.g.,][]{Schonrich11, Beers12}. More studies are needed to
understand the properties of the Milky Way halo. Future surveys, such
a GAIA \citep{Perryman01}, LAMOST \citep[Large Sky Area Multi-Object
  Fiber Spectroscopic Telescope;][]{Newberg09} and LSST \citep[Large
  Synoptic Survey Telescope;][]{Tyson02} will help to shed light on
this matter.

  For M31, halo metallicity studies have probed regions at much
  farther radii than for the Milky Way. \citet{Kalirai06} and
  \citet{Koch08} have detected a clear metallicity gradient over a
  large range of radial distances, from $\sim 10$ kpc to $\sim$ 160
  kpc, with significant scatter around this overall
  trend. \citet{Kalirai06} found a mean metallicity of
  $[\textrm{Fe/H}] \sim -0.5$, $\sim -0.95$, and $\sim -1.26$ dex, at
  average projected distances of 14 kpc, 31 kpc, and 81 kpc from M31's
  center, respectively, from photometric properties of
  spectroscopically selected M31 RGB bulge + halo
  stars. \citet{Koch08} studied a spectroscopic sample of M31's halo
  stars along the minor axis from 9 to 160 kpc and found a strong
  metallicity gradient, with a sharp decline of $\sim 0.5$ dex at
  $\sim 20$ kpc, using the spectral CaT lines. However, several
  studies have found no evidence of such gradient
  ~\citep[e.g.,][]{Durrell04, Irwin05,Chapman06,Richardson09}. Among
  the latest works, \citet{Chapman06} selected a spectroscopic sample
  of M31's halo stars by kinematically isolating nonrotating stars,
  and found a metal-poor metallicity of $[\textrm{Fe/H}] \sim -1.4$
  dex with no detectable metallicity gradient from 10 to 70 kpc. Using
  11 very deep HST fields of M31's halo, \citet{Richardson09} found
  that there is no evidence of a radial metallicity gradient within
  $\sim$30 to 60 kpc. They calculated an average metallicity over that
  range of $[\textrm{Fe/H}] = -0.8 \pm 0.14$ using the photometric
  properties of RGB stars. The differences among the data analyzed and
  methods used to analyze it, as well as the different various halo
  regions probed makes it difficult to completely disentangle why the
  results, mainly in the inner 60 kpc, are so diverse. In general,
  nevertheless, the inner $\sim 15$ to 50 kpc region of M31's halo
  shows a nearly flat, high metallicity profile ($[\textrm{Fe/H}] \sim
  -0.8$ dex), whereas outside $\sim$ 60 kpc the metallicities are
  lower ($[\textrm{Fe/H}] \sim -1.3$ dex).

While the Milky Way and M31 agree qualitatively with the model
predictions of stellar halo formation, they exhibit clear
differences. For instance, the M31 outer halo is more metal rich by a
factor of three or more than that of the Milky Way out to at least 60
kpc \citep{Richardson09}. M31 has had a more active recent accretion
history than that of the Milky Way \citep{Font06b, Fardal06, Brown06,
  Fardal07, Brown08, McConnachie09}. The halo of the Milky Way has a
power-law density profile with exponent $ 2 < \gamma < 4$, and an
estimated oblateness of $0.5 < c/a < 0.8$~\citep[see
  e.g.][]{Newberg_yanny05, Bell08, Juric08}. The halo of M31 has been
characterized with a power-law surface brightness profile of exponent
$\gamma \sim 2$ \citep{Guhathakurta06, Ibata07, Courteau12} and it
contributes $\sim 4\%$ of the total luminosity of M31 out to 200 kpc
along the minor axis \citep{Courteau12}. Given the stochasticity
involved in the process of stellar halo formation, it is important to
enlarge the sample of observed galactic halos to differentiate between
the models and quantify model predictions from galaxy to galaxy.

The GHOSTS (Galaxy Halos, Outer disks, Substructure, Thick disks, and
Star clusters) survey (PI: R. de Jong) is the largest study of
resolved stellar populations in the outer disk and halos of 14 nearby
disk galaxies to
date \footnote{\url{http://archive.stsci.edu/prepds/ghosts}}. A
detailed description of the survey can be found in \citet[][hereafter
  R-S11]{RS11}. Briefly, the targeted galaxies are imaged with the
Advanced Camera for Surveys (ACS) onboard HST and their individual
stars are resolved. GHOSTS observations provide star counts and
color-magnitude diagrams (CMDs) typically 2--3 magnitudes below the
tip of the red giant branch (TRGB) of the outer disk and halo of each
galaxy. Such data allow us to shed light on various issues. For
instance, using the red giant branch (RGB) stars as tracers of the
faint underlying population, we are able to study the size and shape
of each stellar halo. In addition, we can constrain their stellar
properties, such as age and metallicity. These observations can then
be used to help distinguish between models for the formation of spiral
galaxies.

Among the galaxies observed by the GHOSTS survey, M81 (NGC 3031) is of
particular interest. It is the nearest massive spiral galaxy with
similar global properties to the Milky Way and M31, and it contains
the largest number of fields observed within the survey. These
observations probe different regions of M81's halo out to large
galactocentric distances, enabling us to investigate whether stellar
population variations and/or metallicity gradients are found in this
galaxy. The furthest field observed in M81 is at a projected distance
of $\sim 50$ kpc from the galactic center along the minor axis, an
unprecedented distance for halo studies outside the Local Group. M81
thus provides a key laboratory to study the stellar properties of
halos of massive spiral galaxies outside the Local Group and to test
model predictions.

\citet{Tikhonov05} used archival HST/ WFPC2 observations of M81 to
study the spatial distribution of its AGB and RGB stars. They
suggested that the transition from a disk-dominated to a
halo-dominated population occurs at a deprojected distance of $\sim
20$ kpc. Later HST studies of the resolved stellar populations of
fields closer than $\sim 20$ kpc show CMDs exhibiting relatively young
stars and older RGB stars with a wide range in metallicity
\citep{Williams09}. These stars likely belong to the disk of M81.
\citet{Mouhcine05c} have studied the metallicity of RGB stars observed
with a WFPC2 HST field located at a projected distance of $\sim 15$
kpc from the galactic center. They found a mean metallicity of
$[\textrm{Fe/H}] \sim -1.25$ dex assuming an age of 12 Gyr, more
metal-poor than the metallicity inferred from closer fields,
indicating that these are halo stars. \citet[hereafter
    B09]{Barker09} analyzed the outskirts of the northern half of M81
from a wide-field study of resolved stellar populations using the
Suprime-Cam instrument on the 8 m Subaru telescope and covering an
area $\sim 0.3^{\circ}$. They found a flattening of the stellar
density profile beyond a deprojected distance of 18 kpc, (at
approximately 6 scale lengths, being M81's disk scale length $\approx
2.7$ kpc) with stars of metallicities $[\textrm{Fe/H}] \sim -1.1\pm
0.3$ dex, assuming a 10 Gyr old age. B09 discussed the possibility
that this represents the underlying stellar halo of M81.  This outer
component is more metal-poor than the stars found in the interior
region of their field, again suggesting that M81's faint, extended
component starts to dominate at about $R_{dp} \sim 20$ kpc. B09 find
that its RGB stars follow a power-law surface density profile with an
exponent of $\gamma \sim 2$, similar to the halo density profile of
both the Milky Way and M31. They moreover estimate that this component
would contain $\sim 10\%- 15\%$ of M81's total $V$-band luminosity,
several times more luminous than both the MW's and M31's halo,
although the systematic uncertainty in this estimate is about $\pm
50\%$ due to uncertainties in the diffuse light sky subtraction within
14--18 arcmin. The only age estimate for the halo of M81 from its
resolved stellar population was reported by \citet{Durrell10}, who
analyzed very deep observations of one HST/ACS field at $\sim 19$ kpc
from the center of M81. From fits to the red clump (RC), RGB, and RGB
bump, they estimated a mean age for the dominant population of $9\pm
2$ Gyr. They also found that the metallicity distribution function
peaks at $[\textrm{Fe/H}] = -1.15 \pm 0.11$ dex.

  M81 has had an active recent interaction history. HI gas surrounding
  M81, M82 and NGC3077 show filamentary structures
  \citep{Vanderhulst79, Yun94} that can be explained as the result of
  tidal interactions among all three galaxies, probably $\approx
  200-300$ Myr ago \citep{Yun99}. In addition, recent star formation
  activity has been observed in the most prominent HI filaments,
  e.g. Arp's Loop and Holmerg IX \citep{deMello08, Sabbi08, Weisz08,
    Mouhcine09} as well as in the outskirts of M81, especially in the
  HI bridge connecting M81 and NGC3077 \citep{Mouhcine09}. It has been
  suggested that these may be new tidal dwarf galaxies, forming out of
  the gas stripped from the interacting galaxies \citep{deMello08,
    Sabbi08, Weisz08, Mouhcine09}. Using the MegaCam imager on CFHT,
  \citet{Mouhcine09} presented a panoramic view of M81, which covers
  the entire galaxy and the southeast outskirts, including the most
  prominent HI filaments and NGC3077. They found new young stellar
  clumps, whose properties can also be best explained if these systems
  are of tidal origin.

 In this paper, we analyze the stellar populations of M81's halo stars
 by studying the color distribution of a sample of 19 fields near
 M81. We compare our observations with predictions from models of
 stellar halo formation. A companion paper (Vlaji\'c et al., in
 preparation) will present the minor and major axis halo density
 profiles of M81 as traced by the RGB stars observed.  The paper is
 outlined as follows. The observations and photometry are presented in
 Section~\ref{sec:observations}. Analysis of the CMDs and color
 distributions of the M81 fields are in Sections~\ref{sec:cmds}
 and~\ref{sec:cfs}, respectively. Section~\ref{sec:models} introduces
 the method used to transform the model stellar particles into CMDs. A
 comparison of our observations with models of stellar halos formed
 entirely from accreted objects is presented in
 Section~\ref{sec:compa}. We discuss our results in
   Section~\ref{sec:discussion} and conclude in
   Section~\ref{sec:summary}.

All throughout the paper we use the term ``M81's halo'' to refer to
stars located at projected distances $R > 15.5$ kpc from the center of
M81. This is the region where the faint, extended $R^{-2}$ structural
component detected by B09 starts to dominate, at $\approx 6$ disk's
scale lengths. We adopt a distance modulus for M81 of $(m-M)_0=27.79$
(R-S11), which is in agreement with previous values for the distance
modulus of M81, e.g, $(m-M)_0=27.75 \pm 0.08$ \citep{Freedman01},
$(m-M)_0=27.78 \pm 0.05$ \citep{McCommas09}, $(m-M)_0=27.78 \pm 0.04$
\citep{Dalcanton09}, and $(m-M)_0=27.86 \pm 0.06$ \citep{Durrell10}.

\section{Observations and photometry}\label{sec:observations}

\begin{figure}\centering
 \includegraphics[width=80mm,clip]{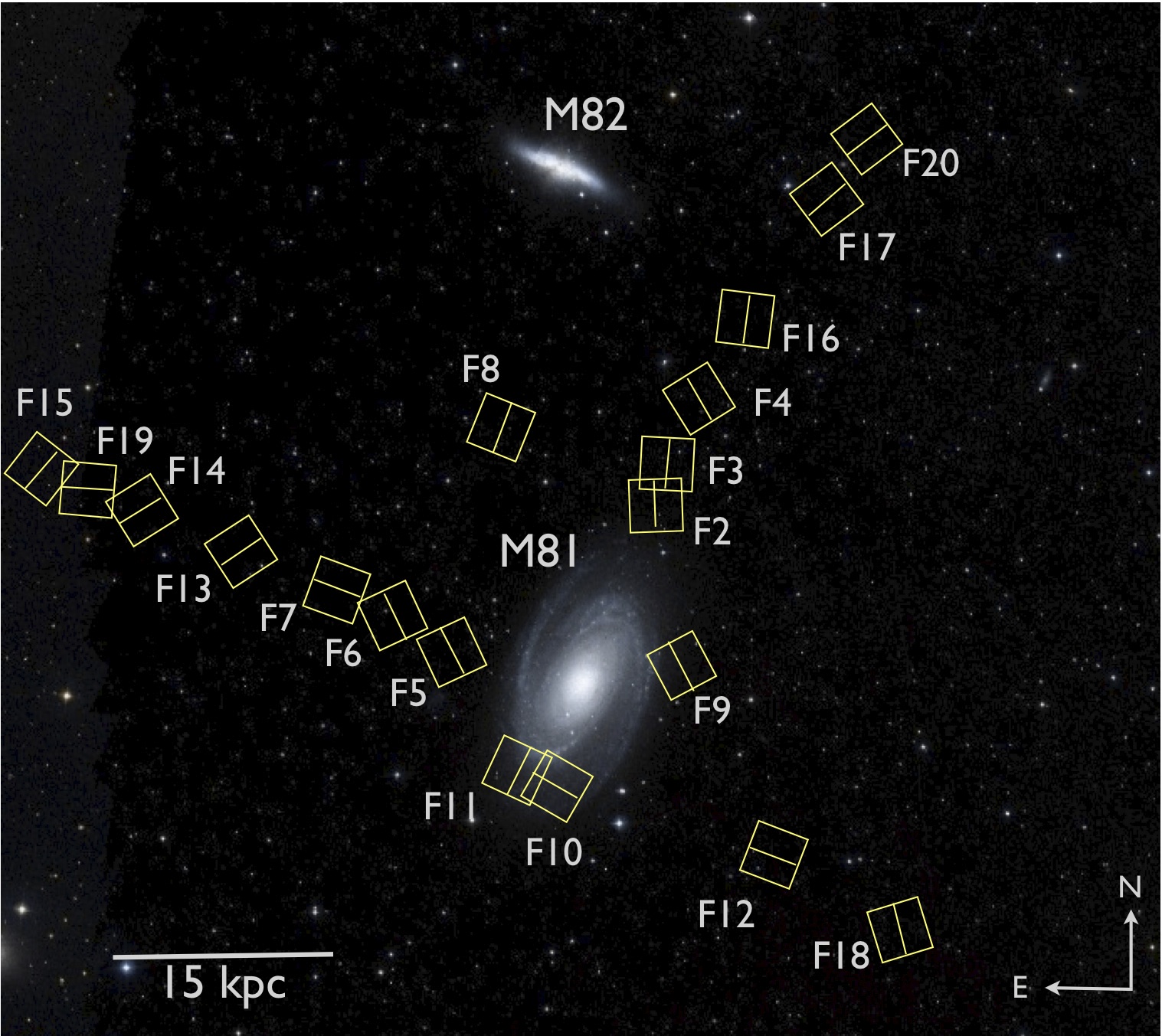}
\caption{DSS colored image of M81 showing the location of the 19
  \emph{HST} ACS/WFC fields used in this work. North is up and East is
  to the left. Fields F2--F12 were already introduced in R-S11 whereas
  fields F13-F20 are new observations. The image is
  $\sim1.25^{\circ}\times1.25^{\circ}$. Each ACS field of view (Fov)
  covers a region of $\sim3.6\times3.6$ kpc$^2$ on the sky at the
  distance of M81 (3.6 Mpc, R-S11), and has a pixel scale of
  $0\arcsec.05$. Information about these observations can be found in
  Table~\ref{table:log}. The satellite galaxy M82 can also be seen.}
\label{fig:location}
\end{figure}

We use \emph{HST} ACS/WFC images in the $F606W$ and $F814W$ filters of
19 fields near M81 from the GHOSTS survey (see R-S11). As shown in
Figure~\ref{fig:location}, most of the pointings are spaced along the
minor and major axes of M81 mainly to measure its halo structure
(Vlaji\'c et al., in prep.). This strategy allows us to probe the halo
of M81 out to a projected distance of $R \sim$ 50 kpc from the
galactic center, an unprecedented distance for halo studies outside
the Local Group. Fields 2--12 were introduced in R-S11, while Fields
13--20 are new observations obtained in our Cycle 17 SNAP program
(Vlaji\'c et al., in prep.). A summary of the GHOSTS fields used in
this work can be found in Table~\ref{table:log}.

\begin{deluxetable*}{lccccccccc}
  \tabletypesize{\scriptsize} \tablecaption{M81 HST/ACS
    observations\label{table:log}} \tablewidth{0pt}
  \tablehead{\colhead{Field\tablenotemark{a}}& \colhead{Proposal ID}&
    \colhead{Proposal PI}& \colhead{$\alpha_{2000}$ (h m s)}&
    \colhead{$\delta_{2000}$ ($^{\circ}$ m s)}& \colhead{Observation
      Date}& \colhead{t\tablenotemark{b}$_{F606W}$ (s)}&
    \colhead{t\tablenotemark{b}$_{F814W}$ (s)}& \colhead{$E(B-V)$
      \tablenotemark{c}} & \colhead{R (kpc) \tablenotemark{d}}}
  \startdata F02 & 10915 & Dalcanton & 09 54 34.71 & 69 16 49.76 &
  2006, Nov 16 & 24232 (10) & 29953 (12) & 0.086 & 15.0\\ F03 & 10523
  & de Jong & 09 54 23.12 & 69 19 56.41 & 2005, Dec 06 & 730 (2) & 730
  (2) & 0.090 & 18.5\\ F04 & 10523 & de Jong & 09 53 59.63 & 69 24
  58.57 & 2005, Oct 26 &695 (2) & 695 (2) & 0.089 & 24.4\\ F05 & 10523
  & de Jong & 09 57 17.23 & 69 06 29.27 & 2005, Oct 31 &700 (2) &700
  (2) & 0.079 & 10.4\\ F06 & 10523 & de Jong & 09 58 04.50 & 69 08
  52.15 & 2005, Oct 31 &700 (2) & 700 (2) & 0.077 & 15.5 \\ F07 &
  10523 & de Jong & 09 58 52.30 & 69 10 42.12 & 2005, Sep 07 &720 (2)
  & 720 (2) & 0.071 & 20.5\\ F08 & 10523 & de Jong & 09 56 39.13 & 69
  22 29.58 & 2005, Dec 20 &740 (2) & 740 (2) & 0.078 & 21.0\\ F09 &
  10136 & Bond & 09 54 16.54 & 69 05 35.71 &2005, Apr 13& 5354 (4) &
  5501 (4) & 0.080 & ~7.6\\ F10 & 10584 & Zezas &09 56 29.23 & 68 54
  42.49 & 2005, Dec 09 & 1580 (3) & 1595 (3) & 0.078 & 11.3\\ F11 &
  10584 & Zezas &09 57 01.91 & 68 55 56.29 & 2005, Dec 06 & 1580 (3) &
  1595 (3) & 0.078 & 12.0\\ F12 & 10604 & Sarajedini & 09 53 03.20 &
  68 52 03.60 &2005, Sep 11&12470 (10)& 22446 (18) & 0.074 &
  19.3\\ F13 & 11613 & de Jong & 10 00 14.30 & 69 13 14.75 & 2010, Jan
  18 &850 (2) & 690 (2) & 0.065 & 29.0 \\ F14 & 11613 & de Jong & 10
  01 34.88 & 69 16 03.11 & 2010, Jul 22 &850 (2) & 690 (2) & 0.062 &
  37.0\\ F15 & 11613 & de Jong & 10 02 56.42 & 69 18 53.08 & 2010, Jan
  23 &850 (2) & 690 (2) & 0.057 & 45.5\\ F16 & 11613 & de Jong & 09 53
  20.88 & 69 30 30.71 & 2010, Jun 03 &850 (2) & 690 (2) & 0.091 &
  31.3\\ F17 & 11613 & de Jong & 09 52 13.33 & 69 38 59.17 & 2010, Jul
  16 &850 (2) & 690 (2) & 0.098 & 42.3\\ F18 & 11613 & de Jong & 09 51
  23.91 & 68 46 29.97 & 2009, Nov 09 &850 (2) & 690 (2) & 0.074 &
  30.4\\ F19 & 11613 & de Jong & 10 02 15.46 & 69 17 33.68 & 2010, Feb
  25 &830 (2)& 680 (2) & 0.059 & 41.4\\ F20 & 11613 & de Jong & 09 51
  41.10 & 69 43 14.85 & 2009, Dec 31 &830 (2) & 680 (2) & 0.094 &
  47.9\\ \enddata \tablenotetext{a} {Field number. Fields 2--12 were
    presented in R-S11 and are labeled following their notation. Fields
    13--20 are new observations from our Cycle 17 program.}
  \tablenotetext{b}{Total time of exposure time. The number of
    exposures in each filter is indicated in brackets.}
  \tablenotetext{c}{Reddening values estimated by
    \citet{Schlegel98}. Note that, as stated in the text, we calculate
    the extinction for each bandpass using the corrected extinction
    ratios by \citet{Schlafly11}, which recalibrate the
    \citet{Schlegel98} dust maps.}  \tablenotetext{d}{Projected
    distance from the galactic center, without taking into account
    M81's inclination.}
\end{deluxetable*}

Stellar photometry was performed using \verb+DOLPHOT+, a modified
version of HST\verb+PHOT+ \citep{Dolphin00} for ACS images. We refer
the reader to R-S11 for full details of the GHOSTS data pipeline
developed for the photometric reduction. Briefly, \verb+DOLPHOT+
performs point-spread function (PSF) fitting on all the flat-fielded
(FLT) images per field simultaneously. The \verb+DOLPHOT+ parameters
used for performing photometry on the GHOSTS fields are similar to
those used in the ANGST program \citep{Dalcanton09}. The final output
of \verb+DOLPHOT+ provides both instrumental VEGA and transformed
Johnson-Cousins magnitudes already corrected for charge transfer
efficiency (CTE) loss and with aperture corrections calculated using
isolated stars. CTE correction is applied by \verb+DOLPHOT+ to
  each source detected using the analytical formulae provided by
  \citet{Chiaberge09}. The photometric output also includes various
diagnostic parameters that can be used to discriminate spurious
detections from the actual stars. Contamination from Galactic
foreground stars was estimated using the Besan\c{c}on model
\citep{Robin03}. We find that only $\sim50$ foreground stars are
expected per ACS field. The most important source of contamination in
these fields are unresolved background galaxies. The estimated galaxy
density was obtained using the GalaxyCount program
\citep{Ellis_blandhawthorn07} and it is 22, 55, 121, and 240
$\rm{arcmin}^{-2}$ at $F814W =$ 24, 25, 26, and 27 mag,
respectively. Several selection criteria to discriminate unresolved
galaxies from stars were optimized using deep archival high-redshift
HST/ACS fields. These were applied to the raw photometric output,
which removed $\sim$95\% of the contaminants. Details on the
photometric culls and how they were optimized can be found in
R-S11. In addition, a mask of all extended and resolved objects was
constructed for each field using SE\verb+XTRACTOR+
\citep{Bertin_arnouts96}. Detections lying in the pixel positions of
the masked sources were discarded from the star catalog.

Extensive artificial star tests (ASTs) were performed in both filters
of all images to assess the completeness level and quantify the
photometric errors of the data.  The procedure of the ASTs are
explained in detailed in R-S11. In short, approximately 2,000,000
artificial stars per field in both filters are injected and
photometered by \verb+DOLPHOT+, one at a time to avoid affecting the
image crowding. The artificial stars were distributed according to the
observed stellar gradient, thus the higher surface brightness regions
of an observation were populated with more artificial stars. The
colors and magnitudes of the injected artificial stars are realistic
and they cover not only the observed values but also fainter
magnitudes to explore the possibility of recovering a faint,
unresolvable star. We applied the same cull as in the real
images. Artificial stars that did not pass the cull were considered as
lost. The completeness level was calculated as the ratio of
recovered-to-injected number of artificial stars at a given color and
magnitude bin. The 50\% completeness level varies from field to field
but it is typically found at $ F814W \sim 26$, which is approximately
2 mag fainter than the tip of the red giant branch (TRGB), measured to
be located at $F814W \sim 23.7$ for M81 (R-S11).  The 50\%
completeness level appears to be rather color independent for most
regions of the CMD. For colors ($F606W-F814W$) $\gtrsim 1.7$, or
equivalently $F606W \gtrsim 27.5$, however, the incompleteness is more
severe (see Figure~\ref{fig:cmds} in the next section).

\section{The color magnitude diagrams of the M81 fields}\label{sec:cmds}

\begin{figure*} \centering
\includegraphics[width=180mm,clip]{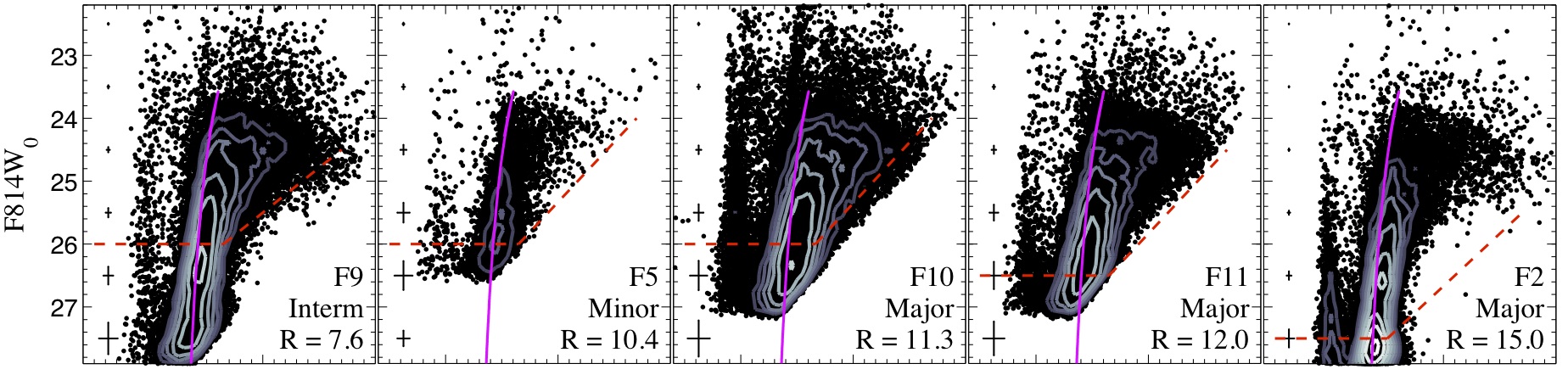}
\includegraphics[width=180mm,clip]{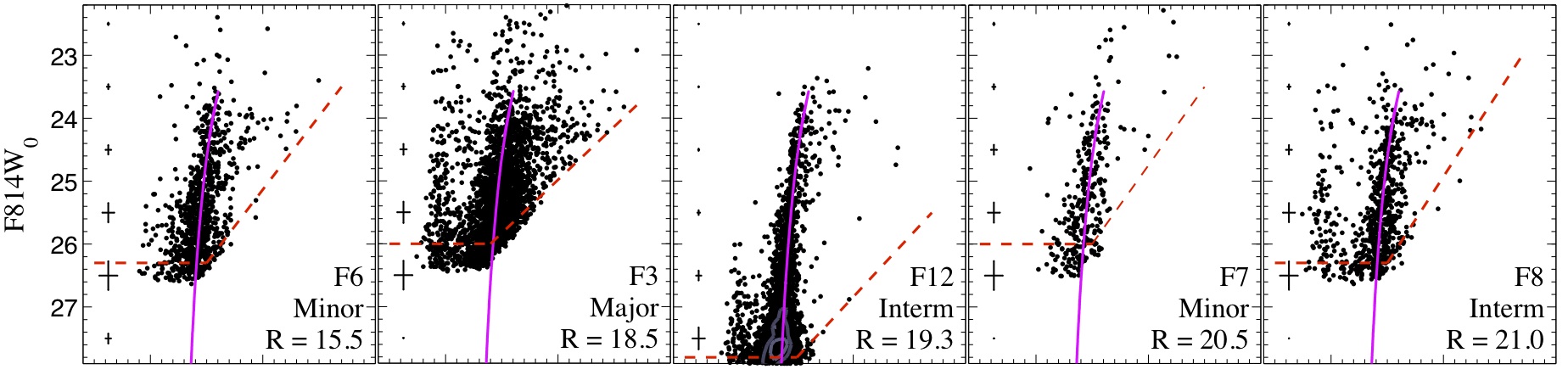}
\includegraphics[width=180mm,clip]{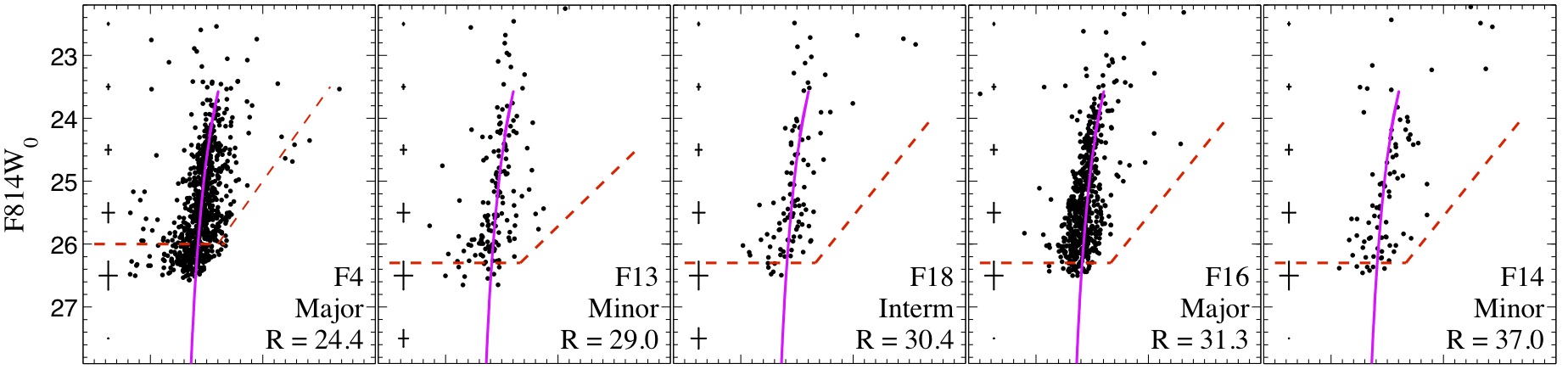}
\includegraphics[width=180mm,clip]{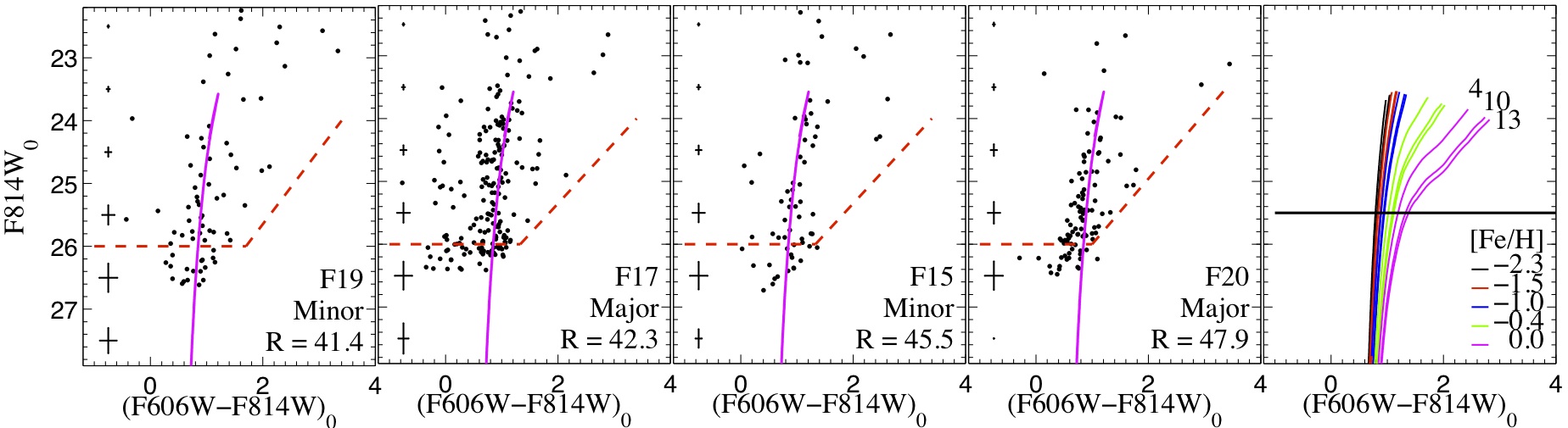}
  \caption{CMDs of the 19 ACS/WFC fields in M81 after correcting for
    extinction and reddening. We show only the stars which remain
    after the culling and masks were applied to the DOLPHOT
    photometric outputs. At densities greater than 40 stars in a
    0.1$\times$0.14 bin in color and $F814W$, a Hess diagram is shown
    with contours at 40, 60, 100, 160, 250, 400, 650, and 1000 stars
    per 0.014 $\textrm{mag}^{2}$. Magnitudes are calibrated onto the
    VEGAmag HST system and corrected for Galactic extinction (see
    Table~\ref{table:log}). Labels ''Major'', ''Minor'', and
    ''Interm'' indicate that the field is located on the major, minor
    or on intermediate axis, respectively. The projected distance, R,
    of each field from the center of M81 is indicated in units of
    kpc. A 10 Gyr old, $[\textrm{Fe/H}]= -1.2$ dex isochrone from the
    BaSTI library \citet{Pietrinferni04} is superposed in each
    panel. Red dashed lines indicate the 50\% completeness level as
    determined from the ASTs.  The photometric errors are also
    obtained from the ASTs and refer only to color ($F606W - F814W$)$
    =1$. The right-most bottom panel shows, with different colors,
    BaSTI isochrones covering a grid of metallicities for 3 different
    ages: 4, 10 and 13 Gyr from left to right of each color which
    indicate the age-metallicity degeneracy present in the RGBs. Old
    and more metal-poor populations resemble younger and more
    metal-rich ones. The black solid horizontal line shows the
    limiting magnitude that we used for the color analysis performed
    in the next sections.}
\label{fig:cmds}
\end{figure*}

Figure~\ref{fig:cmds} shows the color-magnitude diagrams (CMDs) of
stars for each field, after the culling and masks were applied. The
magnitudes have been corrected for Galactic extinction using the
corrected extinction ratios $A_{\lambda}/E(B-V)$ of 2.47 and 1.53 for
$F606W$ and $F814W$, respectively, that are to be used with the E(B-V)
values from \citet{Schlegel98} dust maps and the $R_V=3.1$
  extinction law \citep{Schlafly11}. Given the variable Galactic
extinction across the sampled region of M81, each field was corrected
with its appropriate $E(B-V)$ value as listed in
Table~\ref{table:log}, with a total range of $ 0.05 -
    0.1$. The 50\% completeness level of each field as well as their
projected radial distance from the center of M81 are indicated in each
corresponding panel. The stellar density tends to decrease with
increasing galactocentric radius, as expected. We note that the
projected distances were calculated using circular symmetry, without
taking into account M81's inclination. Therefore, due to the
elliptical symmetry of the disk's density profile, fields located at
similar projected distances but lying on different axes will have
different stellar densities.

The CMDs are mostly populated by old RGB stars (older than 1 Gyr).
Note however that signs of younger populations such as blue, extended
MS stars ( $<$ 500 Myr) or massive stars burning helium in their core
(25--600 Myr old red and blue loop sequence stars) appear primarily in
the fields closer than $R \sim 15$ kpc, which are dominated by disk
stars. In this work, we focus on the stellar populations of M81 fields
located at $R > 15$ kpc ($\approx 6$ disk's scale lengths), where the
faint extended $R^{-2}$ structural component detected by B09 starts to
dominate. A flattening in the surface brightness profile is also
detected using the resolved stars of our GHOSTS fields at $R \sim 18$
kpc (Vlaji\'c et al., in prep.). We detect the faint structural
component discovered by B09 out to a larger projected radius, $R\sim
50$ kpc, and we refer to it hereafter as {\it M81's halo}. Although it
is not yet clear what the nature of this structural component is (see
discussion in B09), we will assume that it is the halo of M81 when
comparing with predictions by models of galaxy formation (see
Section~\ref{sec:compa}). We can gain some insight about the halo's
dominant population by fitting isochrones to the RGBs of their
CMDs. We find that a BaSTI isochrone \citep{Pietrinferni04} of 10 Gyr
with metallicity $[\mathrm{Fe/H}]= -1.2$ dex matches reasonably well
the shape of the RGB, as shown in Figure~\ref{fig:cmds}. This is
consistent with the results by \citet{Durrell10}, who derived a mean
metallicity of $[\mathrm{M/H}] = -1.15 \pm 0.11$ dex and age of $9 \pm
2$ Gyr from the shape of the RGB, the magnitude of the RC, and the
location of the RGB bump of one of the fields (F12) also analyzed in
this work.

We note that fields F14 and F15 have RGBs slightly redder than the
isochrone superposed in Figure~\ref{fig:cmds}. This might suggest that
the halo populations in such fields have metallicities higher than
$[\mathrm{Fe/H}]= -1.2$ dex. In fact, their RGBs can also be well
fitted with a 10 Gyr isochrone of metallicity $[\mathrm{Fe/H}]= -1$
dex. The redder RGBs might indicate the presence of substructure
  dominating these fields which we might be unable to clearly observe
  due to pencil-beam nature of our survey. However, calibration
uncertainties due to, e.g., CTE correction uncertainties between
chips, flat-field \citep{JD12} or different roll angles can account
for up to $\sim 0.05$ mag uncertainty in color. This uncertainty was
determined by comparing the median colors of overlapping stars in
fields F2-F3, F10-F11 in M81 as well as in other fields of the GHOSTS
survey (see R-S11). Therefore, it is unclear whether the difference in
colors are related to metallicity variations in these fields.

On the other hand, fields F8 and F17 have very bright and blue MS
stars at projected distances of 21 kpc and 42.3 kpc respectively,
quite far away from the disk of M81. This is more likely related to
the HI gas that surrounds M81, M82 and NGC3077 and show filamentary
structures \citep{Yun94} that can be explained as a consequence of
close encounters between M81 and its two neighbors about 200--300 Myr
ago \citep{Yun99}. F8 is located in the region of the HI tidal bridge
connecting M81 and M82, called Arp's loop. F17 is also superposed on
HI gas detected in the system. Previous works have suggested that the
bright young stars in the Arp's loop may have formed in the gas
stripped from these interacting galaxies \citep{deMello08} and studies
of a wide-field imaging of the northern half region of M81 show that
the surface density of young stars ($\lesssim 100$ Myr) traces the HI
column density observed (B09). A similar scenario could explain the
presence of young stars in F17. Finally, Field F3 is located at $R >
15$ kpc and it also contains very young main sequence stars, as well
as bright AGB stars above the TRGB. However, since it lies on the
major axis of M81's disk, it is expected to contain a larger
contribution of disk stars than a minor axis field at a similar
projected distance from the galactic center (e.g., F12). In addition,
F3 is located on a high column density of HI as detected by
\citet{Yun94}.

 We also note that fields F16, F17 and F20 are relatively close, in
 projected distance, to M82 (see Figure~\ref{fig:location}). To
 investigate whether there is a significant contribution from M82's
 halo stars in them, we inspected the density distribution of stars in
 the outer areas of this galaxy using ANGST (ACS Nearby Galaxy Survey
 Treasury) data \citep{Dalcanton09}. The data cover the entire disk of
 M82 and to some extent the outer regions, being the outermost field
 at $\sim 3.5$ kpc from M82's center. We find that there is a factor
 of $\sim20$ more stars in M81 than in M82 at a 3.5 kpc projected
 distance from each galaxy. If we assume that the stellar density
 difference remains the same out to larger distances, we can use the
 observed stellar density profile of M81 (see Vlaji\'c et al., in
 prep.) and estimate the corresponding density of M82's stars at $22$
 kpc, which is the projected distance of F20 from M82's center. We
 obtain that the number of M82's stars at 22 kpc is $\sim 1/3$ of what
 is observed in field F20\footnote{We obtain a similar result if we
   consider the distance and stellar density of field F17. Field F16,
   on the other hand, has a factor of 4 more stars than that expected
   from M82's halo.}. We emphasize that this estimate is extremely
 uncertain; a $R^{-2}$-power law fit to the stellar density
 distributions would estimate a factor of $\sim 5$ lower contribution
 from M82 than from M81 at the location of F20. In addition, as we
 show in the next section, we see no differences in colors in the
 fields closest to M82. We therefore assume that the results presented
 below are not significantly influenced by M82's halo.

\section{The Color distribution of M81's halo}\label{sec:cfs}

 As shown in Figure~\ref{fig:cmds}, we mostly observe stars populating
 the RGB in the halo CMDs of M81. Because the age and metallicity are
 partially degenerate in this evolutionary phase \citep[see
   e.g.,][]{Worthey94}, it is impossible to constrain the ages and
 metallicities of the stars from these CMDs alone.  However, the
 effects of age are relatively small compared to metallicity, such
 that the color of the RGB is an approximate indicator of
 metallicity. Here we analyze the colors of the RGBs as a function of
 galactocentric distance, which will then be directly compared with
 predictions by models of galaxy formation (see
 Section~\ref{sec:compa}).

\begin{figure*} \centering
  \includegraphics[width=185mm, clip]{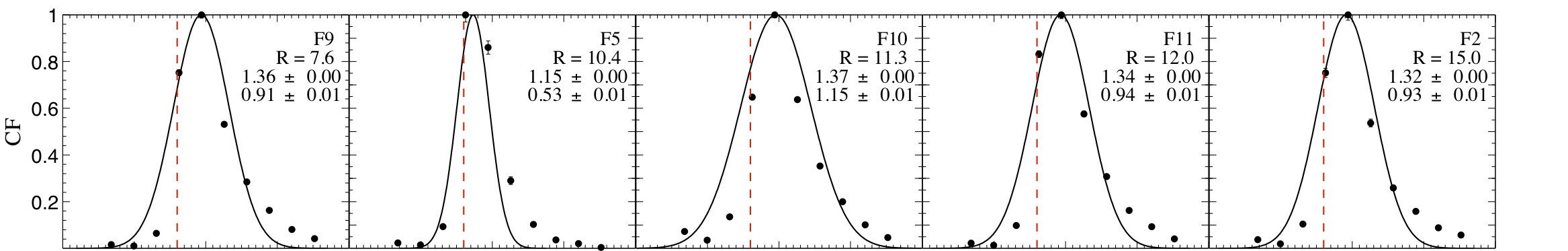}
  \includegraphics[width=185mm, clip]{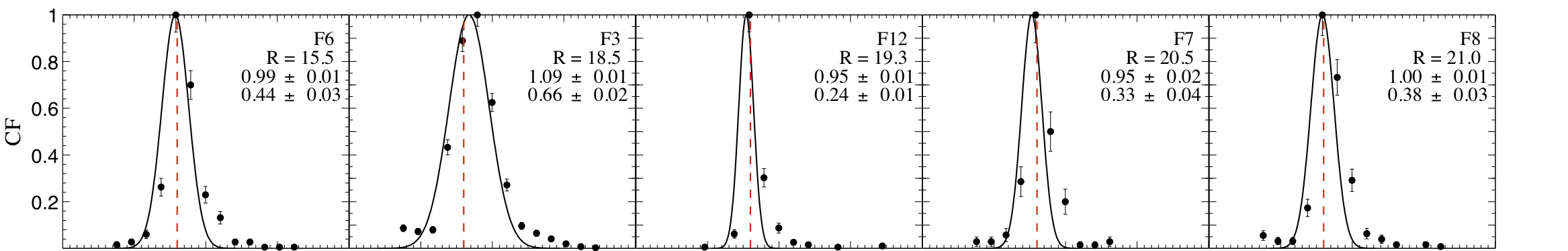}
  \includegraphics[width=185mm, clip]{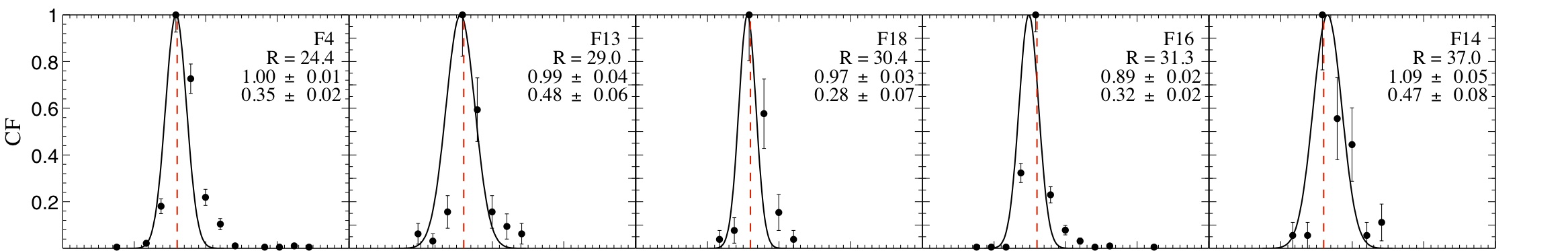}
  \includegraphics[width=185mm, clip]{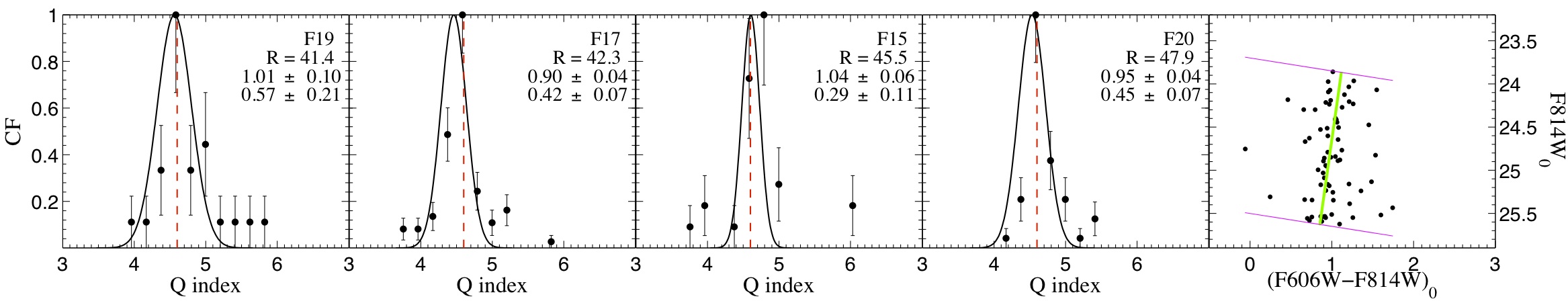}
  \caption{Color function (CF) distribution of each field of M81. Only
    stars within the magnitude range $23.7 \leq F814W \leq 25.5$ were
    considered to build the CFs, as shown in the right-most bottom
    panel. Black dots show the $Q$-color distribution of stars, which
    were divided into 15 bins in the color index $Q$ range $3.5 \leq Q
    \leq 6$, and normalized to one. Error bars to the data points are
    simply Poisson noise. The field number and its projected radial
    distance, R, from M81's center in kpc are indicated. The values
    below the radius correspond to the median of the ($F606W - F814W$)
    color and its corresponding uncertainty, and the width of the CF
    and its uncertainty is written below it. The width of each CF was
    calculated from a Gaussian function, shown here as a solid line,
    which was obtained from fitting the entire data (not the binned
    data) using a maximum likelihood method. Note that the median of
    the color becomes bluer as we get further from the galactic
    center. Also the width appears to mostly become smaller at larger
    galactocentric distances.  The dashed red line in each panel is at
    the same $Q$ value in all the plots to help visualize differences
    in the CFs among the fields. The right-most bottom panel shows the
    CMD of field F13 with a green line superimposed. This line is
    parallel to the isochrone shown in Figure~\ref{fig:cmds} which was
    used as the new $y$-axis to define the $Q$ index. The length of
    the line, as well as the magenta lines, indicate the range of
    magnitudes considered in this analysis.}
 \label{fig:cdf-m81}
 \end{figure*}

 In order to obtain a distribution that better reflects the spread in
 metallicity on a given observed field, we define a new color index
 $Q$ by slightly rotating the CMDs an angle of $-8^{\circ}.29$, where
 a line of slope $-6.7$ becomes vertical. The rotation is such that
 the magnitude axis ($y$-axis) of each CMD is parallel to the
 $[\mathrm{Fe/H}]= -1.2$ dex isochrone shown in Figure~\ref{fig:cmds},
 which represents well the RGB of the halo
 fields. Figure~\ref{fig:cdf-m81} shows the normalized $Q$ color
 function (CF) distributions for each field, plotted as black dots in
 bins of $\Delta Q = 0.3$. The error bars simply indicate Poisson
 noise. The field number and the projected radial distance from the
 center of M81 in kpc are indicated in each panel. The CFs are
 calculated using stars within the magnitude range 23.7 $\leq F814W
 \leq 25.5$. The lower magnitude limit ensures that stars are brighter
 than the 50\% completeness level in all the fields and have small
 photometric errors, while the upper magnitude limit corresponds to
 approximately the TRGB magnitude and thus minimizes contamination
 from bright AGB stars or other contaminants. The right-most bottom
 panel in Figure~\ref{fig:cdf-m81} shows both the line parallel to the
 isochrone (green line) as well as the magnitude range considered
 (within the magenta lines) to construct the CFs for one of the
 observed CMDs (field F13).

 The widths of the CFs, which provide an idea of the range in colors
 at any given radius, vary from field to field. The CFs become
 generally narrower as we get further out from the galactic center.
 To quantify the CF widths, we fitted a Gaussian function to the $Q$
 color distribution of each field using a maximum likelihood
 method\footnote{Note that the parameters of the Gaussian function are
   estimated from all the data points and not from the binned
   data.}. We iteratively rejected $3\sigma$ outliers to avoid fitting
 the tails of the distributions. Since the CMDs have different depths,
 we took into account the photometric errors on the individual data
 points when fitting the Gaussians, in order to avoid introducing
 systematic errors on the CF widths. The resultant intrinsic CF widths
 were calculated as the FWHM of each Gaussian distribution, plotted as
 a solid line in Figure~\ref{fig:cdf-m81}. The median ($F606W-F814W$)
 colors, obtained by rotating the median $Q$ to the original axes, as
 well as the CF widths are indicated in Figure~\ref{fig:cdf-m81}. The
 median ($F606W-F814W$) colors are associated with the median magnitudes
 of the stars used in the Gaussian fits. Uncertainties in the maximum
 likelihood estimates of both the median colors and widths are also
 indicated and take into account Poisson counting uncertainties and
 photometric errors. As shown in Figure~\ref{fig:cmds}, the
   photometric incompleteness affects the reddest regions of the CMDs
   ($F606W - F814W \gtrsim 1.8$) almost exclusively in the inner
   fields. However, the number of stars in the red end is rather low
   compared with the number of stars in the entire CMD, therefore we
   do not expect this to be a big concern in our analysis. At most, a
   correction for this effect would yield to slightly redder median
   colors for the inner fields.

\begin{figure} \centering
\includegraphics[width=90mm,clip]{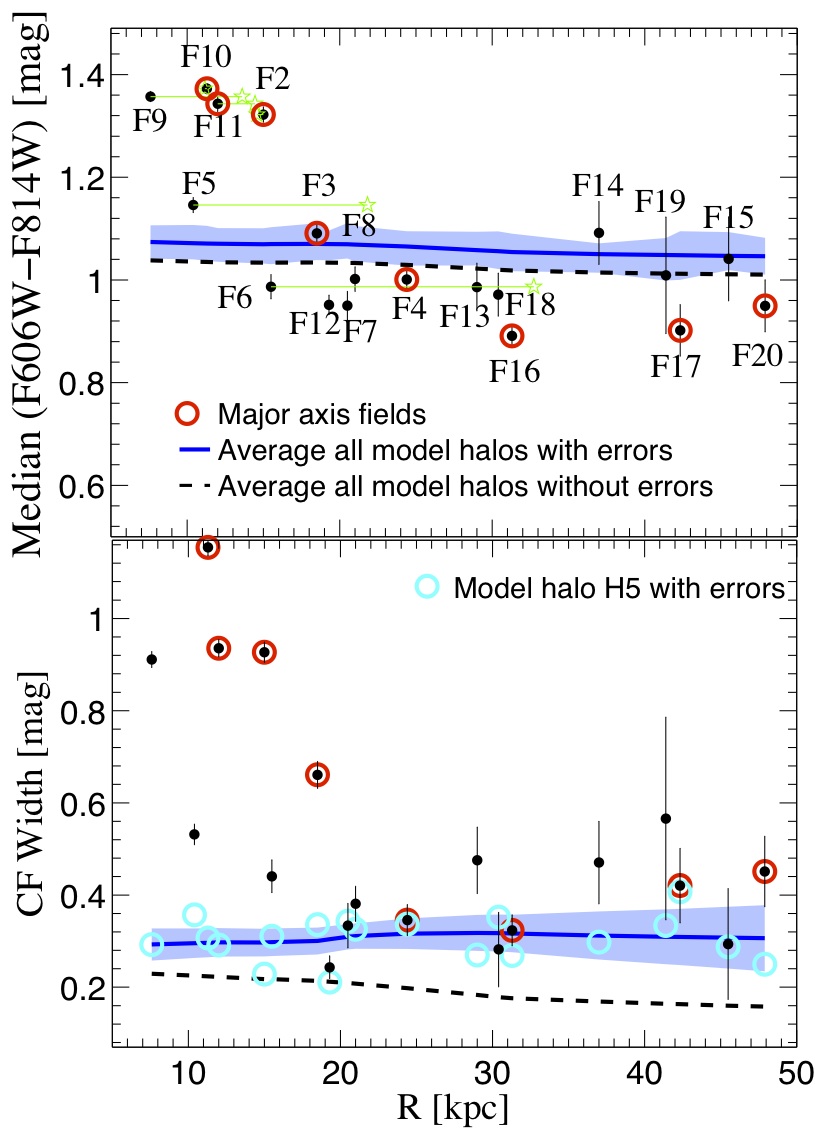}
  \caption{{\it Top panel}: Median color profiles. Black dots indicate
    this value for the M81 observations as a function of projected
    distances $R_{proj}$ to the galactic center. The error bars
    indicate the median uncertainty. Red circles around black dots
    highlight fields located on the major axis of M81. Green stars
    indicate the deprojected distances of fields within $R < 15.5$
    kpc. The colors of these fields (F2, F5, F6, F9, F10, F11)
    decrease with increasing major-axis distance (see text), which
    supports the idea that these fields are mostly dominated by disk
    stars. Thus, assuming that fields located at projected distances
    $R > 15.5$ kpc are pure halo populations, \emph{we do no detect
      any color gradient in the halo of M81}.  The average color
    profile of the 11 smooth halo components of the models analyzed
    (see Section~\ref{sec:models}), is shown as a blue solid line and
    the shaded area indicates the 1--sigma model-to-model deviations
    from the average. Both observations and models lack a color
    gradient. The dashed black line is the average color profile of
    the model fields, before the observational effects were simulated
    as explained in Section~\ref{sec:models}. {\it Bottom panel}:
    Widths of the CFs as a function of projected galactocentric
    distance. The observed M81's fields show a broader CF at $R < 15$
    kpc and a smaller range in colors for fields further out. The
    widths obtained for the models agree with those for observed
    fields at $R>15$ kpc. Again the dashed black line shows the
    average widths for the model CFs before the observational effects
    were simulated. Cyan open circles show the widths as a function of
    radius for one halo model, H5. See text for more details.}
 \label{fig:mod-data1}
 \end{figure}

These results are better illustrated in
Figure~\ref{fig:mod-data1}. The top panel shows the median
  ($F606W - F814W$) colors as a function of galactocentric distance.
Black dots represent the values obtained for the observed CMDs. The
error bars show the median color uncertainties. Red circles around
black dots highlight fields lying on the major axis of M81. There is a
sharp jump in the observed colors at $R \sim 15.5$ kpc.  Fields
closer than $R\sim$15.5 kpc from the center of M81 show a
significantly redder ($F606W - F814W$) color, $\sim 1.35$ mag, than
those located at further distances, color $\sim 1.0$ mag. One
exception is F5, which is closer than 15 kpc but exhibits a color
similar to that of fields farther out. Since F5 is located on the
minor axis of M81, its populations are less disk-dominated than fields
whose projected distances are similar but lie on the major axis of
M81, e.g. F10 and F11.

 In general, the jump in the observed colors is likely due to the
 contribution from disk stars, which becomes negligible outside $R =
 15.5$ kpc, although we cannot rule out the possibility that some of
 the stars at $R < 15.5$ kpc might have higher scale height than the
 disk\footnote{Observations of more edge on galaxies will help to
   understand better halo flattening, and how it changes with
   radius.}. To explore the nature of the stars in fields at $R < 15$
 kpc, we use the ellipticity of M81's disk, assuming an inclination of
 $62.7^{\circ}$ and a position angle of $157^{\circ}$, as listed in
 HyperLeda\footnote{\url{http://leda.univ-lyon1.fr/}}~\citep{Paturel03},
 to generate ellipses passing through the inner fields and calculate
 their deprojected distances. Green stars in the top panel of
 Figure~\ref{fig:mod-data1} indicate those ellipses' semi-major axes
 length. The colors of these fields (F2, F5, F6, F9, F10, F11)
 decrease with increasing semi-major axis distance, which is
 consistent with the idea that they follow the disk's ellipticity and
 thus are mostly dominated by disk stars.

Assuming that fields at projected distances $R > 15$ kpc are pure halo
populations, i.e. belonging to the extended $R^{-2}$ structural
component detected by B09 as well as in the GHOSTS fields (Vlaji\'c et
al., in preparation), an interesting result from
Figure~\ref{fig:mod-data1} is that \emph{no sign of a color gradient
  is observed in the halo of M81.}  We detect, nevertheless, a small
degree of scatter in the median color of 0.055 mag root-mean-square
(RMS). However, as briefly mentioned in section~\ref{sec:cmds}, there
are uncertainties due to the star's image position that are not
reflected in the photometric uncertainties, both from DOLPHOT and
ASTs. These uncertainties are possibly due to CTE corrections between
chips, errors in the flat-field \citep{JD12} or differences in the
roll angle, and have been tested using overlapping stars of three
HST/ACS fields within the GHOSTS survey (see R-S11). The mean
instrumental uncertainty in ($F606W - F814W$) colors is $\sim$ 0.05
mag. Thus, the level of variation detected is consistent with being
due to (mainly) systematic and random errors. Therefore, we are
incapable of distinguishing whether the scatter in color is physical,
indicating real age/metallicity variations, or instrumental. On the
other hand, if we consider the possibility of a color gradient in
M81's halo, a linear fit to the colors of fields farther than 15 kpc
from M81's center yields a slope (or color gradient) of
$-0.0009\pm0.0031$ mag/kpc with a significant scatter around the
median color--distance relation, with $\sigma\sim 0.056$ mag. This is
consistent with no color gradient. Using the obtained slope, we can
place a limit of $0.03\pm0.11$ mag difference between the median color
of RGB halo stars at $\sim$15 and at 50 kpc, corresponding to a
difference of $0.08\pm0.35$ dex in [Fe/H] over that radial range, for
an assumed age of 10 Gyr.

 The bottom panel of Figure~\ref{fig:mod-data1} shows the CF widths as
 a function of galactocentric distance. We find that there is also a
 jump in the observed widths, such that fields closer than 15 kpc have
 broader CFs than those located further away. The uncertainties in the
 widths are larger at increasing radius, due to the small number of
 stars found at large galactocentric distances. The color distribution
 width of $\sim 0.4$ mag for the fields at $R > 15$ kpc suggests that
 the halo fields possess a spread in metallicity, in spite of the
 observational effects. We discuss the possible spread in metallicity
 and return to this figure in general later in the next section, when
 we compare these results with halo model predictions. We have also
 measured the color profile using only stars closer to the TRGB,
 within the magnitude range 23.6 $\leq F814W \leq 24.3$, where there
 is a more sensitive color variation, but much fewer stars. The
 results that we obtain remain overall the same, except that the
 median colors of each field become redder. In addition, the jump
 observed between the colors at $R\sim 15$ kpc increases when
 considering only stars near the TRGB, as the median colors of fields
 at R < 15 change by $\sim 0.35$ mag with respect to the colors shown
 in Figure~\ref{fig:mod-data1} whereas this change for fields at
 larger radii is smaller, $\sim 0.15$ mag.

 We note that the median colors of the RGB stars in regions where the
 HI filaments have been found \citep{Yun94}, mainly fields F3, F5, and
 F8, are not peculiar with respect to the ones elsewhere in the
 outskirts of M81. These fields do contain main sequence stars,
 i.e. evidence of younger populations, in comparison with fields at
 similar distances but lying outside the HI filaments. Our results
 thus suggest that the recent interactions among M81, M82, and NGC3077
 has little influence on the RGB old stars distribution.  On the other
 hand, fields F17 and F20 might have contributions from M82 and the
 determination of their median colors might be more uncertain. They
 are located at $\sim 20$ kpc from M82 whereas their distance from M81
 is $\sim 45$ kpc. However, M81 is a much larger and massive galaxy
 and therefore its contribution to fields F17 and F20 is expected to
 be more than that of M82. We estimated that the M82's contribution
 could be $\sim1/3$ in these fields, but this is highly uncertain (see
 last paragraph of Section~\ref{sec:cmds}.)

In summary, we can characterize the RGB of the halo of M81 by a color
distribution of width $\sim 0.4$ mag and approximately constant median
value of ($F606W - F814W$) $\sim 1$ mag with variations of 0.055 mag
RMS, over a range of $15 \leq R \leq 50$ kpc. The intrinsic color
variation is $<$ 0.055 mag, which is consistent with being due to
random and systematic error alone. When considering the possibility of
a color gradient, we find a limit of $0.03\pm0.11$ mag difference
between the median color of RGB halo stars at $\sim$15 and at 50 kpc,
which corresponds to a difference of $0.08\pm0.35$ dex in [Fe/H] over
that radial range.

\section{Stellar halo models: from star particles to CMDs}
\label{sec:models} 

To compare the properties of the observed stars with models of galaxy
formation, we use the set of stellar halo models presented by
\citet[][hereafter BJ05]{BJ05} and construct CMDs of ACS-like
fields. These stellar halos are formed by the pure accretion of
satellite galaxies within a $\Lambda$CDM cosmology. Full details about
these cosmologically motivated simulations can be found in previous
works \citep[BJ05;][]{Robertson05, Font06}. Here we provide a brief
summary of their main characteristics.

To determine the accretion history of each galactic halo, a merger
tree is generated using the extended Press-Schechter formalism
\citep{Lacey_cole93}. A self-consistent $N$-body simulation follows
the dynamical evolution of the dark matter component of the accreted
satellites, which are being disrupted within an analytic,
time-dependent halo$+$disk potential. A cosmologically-motivated
semianalytic formalism is used to both follow the gas accretion
history of each satellite and model its star formation rate. Star
formation is truncated soon after each satellite falls into the main
halo potential, when the gas is assumed to be lost due to ram pressure
stripping \citep[see e.g.,][]{Lin_faber83, Mayer06}.  The stellar
components of each satellite are associated with the more tightly
bound dark matter particles in the halo, which are assigned a
radially-dependent mass-to-light ratio that produces a reasonable
light profile for the model satellites. The chemical evolution of each
satellite is modeled with the method of \citet{Robertson05}, which
takes into account the enrichment from both Type II and Type I
supernovae.

 BJ05 performed 11 halo realizations. They reproduce several observed
 properties of the Milky Way surviving satellites, such as their
 luminosity function, the luminosity-velocity dispersion relation, and
 their surface brightness distribution. The final stellar halo
 luminosities are comparable to the estimated total luminosity for the
 stellar halo of the Milky Way. It is worth noting that the
 luminosities as well as other properties of the stellar halo
 realizations span a certain range such that these models are also
 suitable for comparing the halo properties of M81, whose luminosity
 and mass are similar to those of the Milky Way \citep[see
   e.g.][]{Karachentsev05}. Differences among the 11 realizations are
 purely due to differences in the model accretion history. The
 publicly available outputs of the
 simulations\footnote{\url{http://www.astro.columbia.edu/~kvj/halos/}}
 provide, among other things, ages, metallicities and masses of the
 stellar populations associated with particles that make up a Milky
 Way-like halo at redshift $z=0$. For the analysis below, we only
 consider the smooth component of each stellar halo model, thus
 neglecting stellar populations that belong to surviving satellites,
 i.e. stellar particles that are still gravitationally bound to their
 original progenitor (see BJ05).  We refer to each halo model as H1,
 H2,..., H11.

We rotate the models by $60\,^{\circ}$ around the $X$ axis, to
simulate M81's inclination, and generate ACS-like fields with
particles located within 5 to 50 kpc, resembling the locations of our
GHOSTS fields. For each of the mock fields, we construct a stellar
``mock-CMD'' containing the predicted mixture of stellar
  populations as follows.

\begin{itemize}

\item We first grid the total range of ages and metallicities
  available in the full halo model ($2\leq$age$\leq 13$ Gyr and $-4
  \leq [\textrm{Fe/H}] \leq -0.4$ dex) with a regular mesh of bin size
  = 1 Gyr$~\times~$0.2 dex.

\item For each mock field, we assign particles of given ages and
  [Fe/H] into the corresponding bins of the grid. We compute the total
  stellar mass associated to each bin, or single stellar population
  (SSP), as the sum of the particle's masses assigned to it.

\item For each age--[Fe/H] bin, we generate a synthetic SSP--CMD using
  IAC-STAR code \citep{Aparicio_gallart04}, assuming a constant star
  formation rate and a uniform metallicity distribution. We adopted
  BaSTI stellar library \citep{Pietrinferni04} and a \citet{Kroupa02}
  initial mass function (IMF). We used the bolometric corrections by
  \citet{Bedin05} to transform the theoretical tracks into the ACS/WFC
  photometric system.

\item Given the total mass associated with each bin, we need to
  calculate the number of stars that the corresponding SSP-CMD is
  contributing to the final mock-CMD. Note that the IAC-STAR code
  takes as input the number of stars rather than the mass of the
  simulated stellar system. Therefore, to realistically convert the
  BJ05 models into synthetic CMDs we need to calculate the
  corresponding number of stars using the stellar mass provided by the
  models. This was estimated using the luminosity functions (LFs)
  provided by Padova web interface CMD code \citep{Marigo08,
    Girardi10}. For a given SSP, the code provides a LF normalized to
  the initial stellar mass of the population.  We multiply this LF by
  the total stellar mass associated with the SSP, which sets the
  number of stars needed.

\item Lastly, in order to make a fair comparison between the
  properties of observed stars and those predicted by the models, we
  simulate the observational effects (incompleteness and photometric
  errors) in the mock-CMDs. We follow the procedure described by
  \citet[][and references therein]{Hidalgo11} for which each star in
  the mock-CMD is applied a magnitude and color correction from the
  AST results. The correction is the difference between the injected
  and recovered magnitudes of a randomly selected artificial star with
  similar injected magnitude, color, and position than the star in the
  mock-CMD.

\end{itemize}

We note that the simulated halos contain a non-negligible fraction of
old ($> 10$ Gyr) metal-poor stars ($[\textrm{Fe/H}] < -2.2$
dex). However, the stellar evolutionary models, and thus the isochrone
data base, do not contain stars with such low metallicities. We have
therefore assigned a metallicity of $[\textrm{Fe/H}] = -2.2$ dex to
all the particles with metallicities lower than that value. Given the
weak dependance of isochrone color on metallicity in this very low
metallicity regime (see Figure~\ref{fig:cmds}), this assumption should
have little impact on the modeled-CMD.

\begin{figure*} %\centering
  \includegraphics[width=57mm,clip]{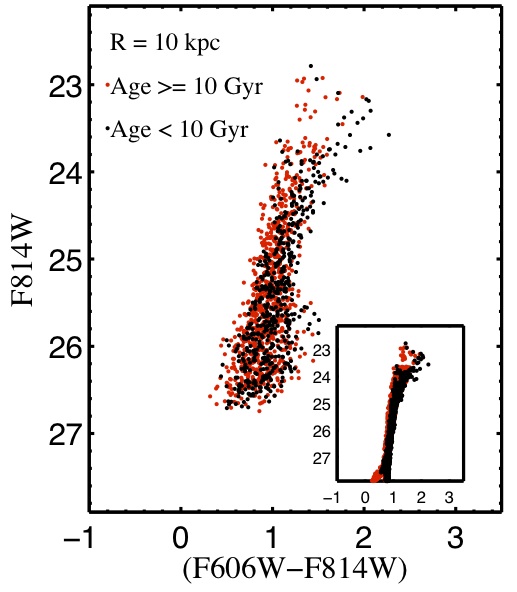}
 \includegraphics[width=57mm,clip]{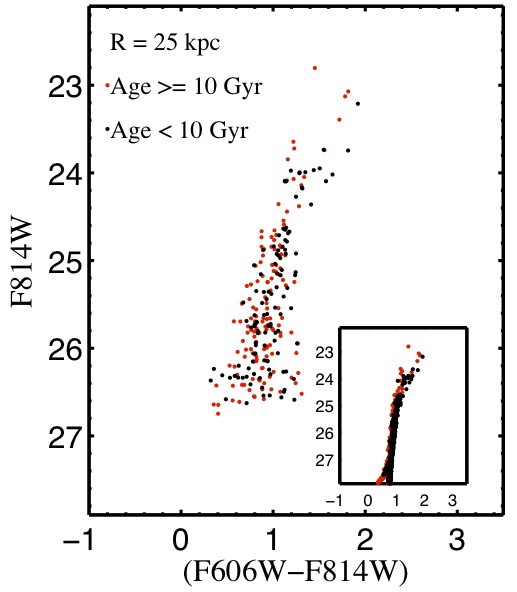}
 \includegraphics[width=57mm,clip]{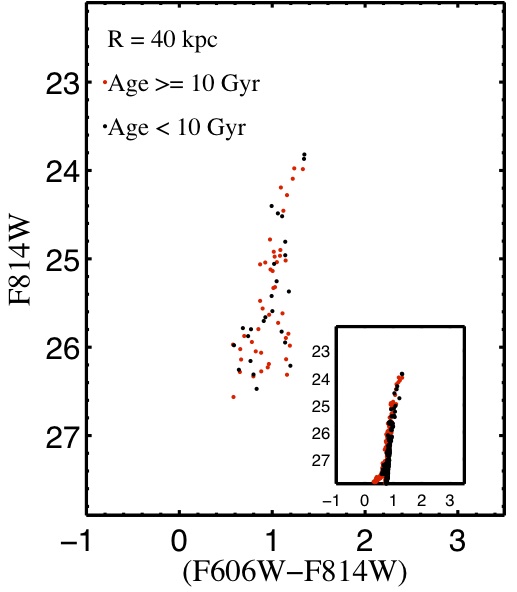} 
\centering
\includegraphics[width=88mm,clip]{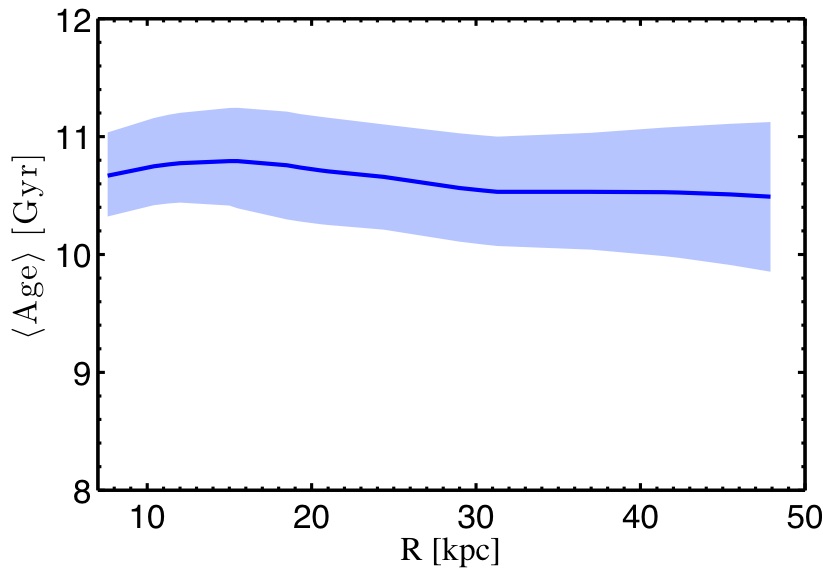}
%%%%%\hspace{0.5cm}
\includegraphics[width=90mm,clip]{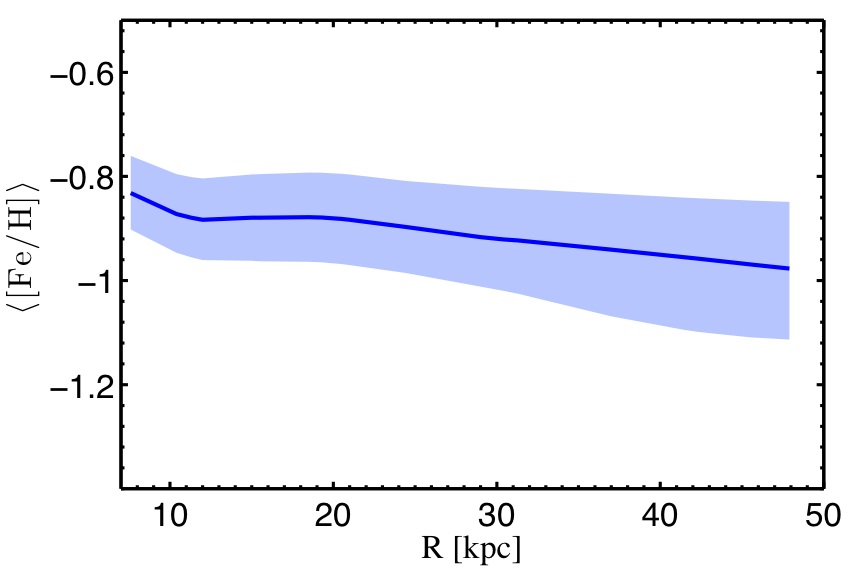}
\caption{{\it Top panel}: Three model CMDs of ACS-like fields
  generated from the stellar-halo model H1 by \citet{BJ05} at
  different galactocentric distances as indicated in each
  panel. Observational effects were simulated using the information
  provided by the ASTs, as described in the text. The insets show the
  model CMDs before simulating the observational effects. Red dots
  represent stars older than or equal to 10 Gyr, whereas black dots
  represent stars with ages between 7 and 10 Gyr. The first feature
  that we notice from these CMDs is their narrow RGBs, despite the
  wide range of ages and metallicities predicted by the models (ages
  from $\sim 6$ to $\sim13.5$ Gyr and metallicities $-4 <
  [\textrm{Fe/H}] < -0.4$ for the particular halo model used
  here). Also, the younger stars preferentially exhibit redder colors
  than the older ones, which is expected due to their higher
  metallicities. {\it Bottom panel}: Luminosity-weighted mean ages
  (left) and [Fe/H] (right) of the constructed CMDs. The solid lines
  indicate the mean ages and metallicities after averaging the results
  obtained for all 44 model fields, which accounts for the 11 halos
  placed at four different orthogonal directions. The shaded area
  indicates the 1--sigma model-to-model scatter from those means. The
  mean age remains roughly constant as a function of radius, whereas
  the mean metallicity becomes slightly more metal-poor at larger
  galactocentric distances, given the inside-out growth of the stellar
  halos assumed for the models.}
 \label{fig:scmds}
 \end{figure*}

The top panel of Figure~\ref{fig:scmds} shows three randomly chosen
CMDs built using the first of the BJ05 halo realizations, H1, at
distances of 10, 25, and 40 kpc from the galactic center. The insets
show the corresponding model CMDs before the observational effects
were simulated. Red dots indicate stars with ages older than or equal
to 10 Gyr, whereas black dots indicate stars with ages between $\sim$7
and 10 Gyr. Despite the significant mixture of populations predicted
by the models, we notice that the RGBs appear to be quite narrow
(especially before simulating the observational effects). We find that
this is always the case, regardless of the model that is used. This
tightness results, at least partly, from the age-metallicity
degeneracy. The younger stars, which should be bluer, are also more
metal rich and therefore have redder colors, which preserves the
tightness of the RGB.

The bottom panel of Figure~\ref{fig:scmds} plots the
luminosity-weighted mean ages and metallicities of the model fields as
a function of galactocentric distance. The solid lines indicate the
average over all 11 halos, after placing the model fields at 4
different orthogonal directions in each halo, giving 44 values
averaged in total. The shaded area represents the 1--sigma
model-to-model scatter of those means.

The mean luminosity-weighted age is rather constant with radius, at
$\sim$ 10.6 Gyr, although there seems to be a mild decrease at larger
radii. The average metallicity, however, shows a weak negative slope
such that the mean metallicity is $\sim 0.12$ dex more
metal-poor at $R\sim 50$ kpc. This gradient is a consequence of the
modeled halo's merger history. The inner regions are assembled earlier
from a few massive more metal-rich satellites, whereas satellites
accreted at later times, which had preferentially more time to form
stars and therefore contain younger stars, populate the halos at
larger galactocentric distances \citep[][]{BJ05, Font06,
  Font08}. \citet{BJ05} have shown that although the recent events
represent a sub-dominant fraction of the total stellar halo luminosity
(from $\sim 5$\% to $\sim50$\%), they become the dominant contributor
at radii of 30-60 kpc and beyond. Thus, the halos at larger radii will
have on average more metal-poor and somewhat younger populations.

\section{Comparing observations with models} 
\label{sec:compa}

We generate CFs and Gaussian distributions for the model-CMDs exactly
as we did for the observed ones, i.e., using the magnitude range and
color index $Q$ described in Section~\ref{sec:cfs}. We calculate the
median colors in each of the model CMDs as well as the width of their
CF distributions. We find that {\it the models do not predict a color
  gradient in the stellar halo}. Instead the color distribution has a
fairly constant median value of ($F606W-F814W$) $\sim 1.1$ at all
galactocentric radii out to 50 kpc, as shown by the blue solid line in
the top panel of Figure~\ref{fig:mod-data1}. This line shows the color
profile of a typical stellar halo, constructed by averaging over all
11 halos at 4 orthogonal positions as explained in
Section~\ref{sec:models}. Nevertheless, the models predict a small
degree of scatter from model to model at a given radius, as shown by
the the shaded area which indicates the 1--sigma color deviation from
the average value. This result is supported by our M81 observations
for fields located at distances larger than 15 kpc, where we detect no
color gradient and an approximately constant value with a small degree
of variation from radius to radius.

We note that the observed median color of M81 appears to be slightly
bluer than the averaged color predicted by the models. It is likely
that this difference is due to oversimplifications in modeling the
satellite galaxy formation and enrichment histories, coupled with
differences in the assembly history of M81's halo with respect to the
Milky Way-like halo modeled by BJ05. In addition, it is possible that
the metallicity floor adopted in the isochrones of $[\textrm{Fe/H}] =
-2.2$ may also be partly responsible for this offset. In spite of the
small mean color offset, however, it is important to emphasize that
the models and observations do agree in the lack of a halo color
gradient. The black dashed line in Figure~\ref{fig:mod-data1}
indicates the color profile, averaged over all modeled fields before
simulating the observational effects. Comparing this line with the
blue solid line shows that the average modeled colors become
systematically redder once the observational effects are taken into
account. This shift is likely due to the larger photometric errors of
redder RGB stars compared to the bluer stars in the same
$F814W$-magnitude range (the latter with brighter $F606W$ magnitudes).

The bottom panel of Figure~\ref{fig:mod-data1} shows the comparison
between the observed and modeled widths of the CFs as a function of
galactocentric distance. The solid line and shaded areas indicate, as
in the top panel, the average width and the 1--sigma model-to-model
scatter for the model halos, respectively. The widths obtained from
the modeled CFs are in good agreement with the observed values for
fields located further than 15 kpc from M81's center; interior to
this, the observed CMDs are likely to have a substantial disk
component, which is not included in the BJ05 halo models. The black
dashed line indicates the average CF's width for the modeled fields
before the observational effects were simulated, which is $\sim 0.2$
mag. A color distribution width of $\sim 0.2$ mag corresponds to a
spread in metallicity of the order of $\sim 0.6$ dex, as predicted by
the halo models. The widths of the CFs are clearly larger after the
observational effects are applied due to the photometric errors which
broaden the RGBs. We note that at smaller radii the observational
effects do not appear to significantly affect the CF widths. This is
because the exposure time of fields, e.g., F9 and F2, which are at
small radii, are much larger than the exposure time of most fields and
therefore their photometric uncertainty are smaller. The cyan circles
show the CF widths of one of the halo models after simulating the
observational effects. The cyan symbols lying on the black dashed line
(i.e. which represents the modeled widths before the observational
effects were simulated) are associated with the locations of the
observed fields that have the largest number of exposures (fields F2,
F9, and F12; see Table~\ref{table:log}). Therefore the broadening
caused by observational errors on the widths of these particular CFs
is almost negligible. In general, we find that the observed widths can
be remarkably well reproduced by the models once the observational
effects are taken into account. Given that the CF widths of the
  synthetic CMDs before the observational effects are simulated
  correspond to a metallicity spread, the good agreement between the
  observed and model widths, once the observational effects are taken
  into account, suggests that there may be a similar spread in
  metallicity ($\sim 0.6$ dex) in the observed M81 halo fields, even
  though the color distributions widen due to observational effects.

As discussed before, no color gradient is found in both the observed
and modeled halos. However, within the hierarchical paradigm, the
outer regions of halos ($R > 20$ kpc) are expected to have a
noticeable spread in the ages and metallicities of their stellar
populations \citep[see, e.g.,][]{Font08}, mainly due to the presence
of substructure in the form of cold stellar streams. Panoramic views
of M31 \citep{Ibata07, McConnachie09} have indicated that the presence
of significant variations in the halo stellar populations can be
associated with observed substructures \citep[see e.g.,][]{Brown06,
  Richardson08}. Bearing in mind that the analysis in this work is
based on pencil-beam HST observations, which sparsely sample M81's
halo, we ask the following question. How likely is it that signatures
of the substructure predicted by simulations, such as variations in
the mean color of the stellar populations, are rendered difficult to
observe due to the pencil-beam nature of these observations?

\begin{figure*} \centering
\includegraphics[width=180mm,clip]{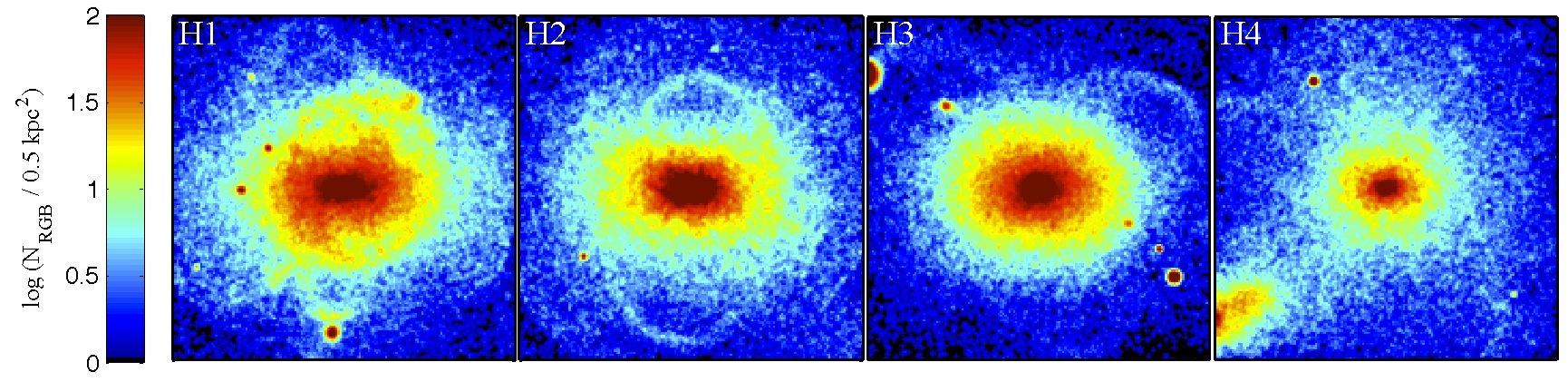}
%%%%%\hspace{0.5cm}
\caption{Density maps of modeled RGB stars, at the distance of M81,
  and rotated according to M81's inclination. Four different halo
  realizations were used, H1, H2, H3, and H4. The maps are shown in a
  $100\times100$ kpc$^2$ area, on the $XZ$ projection. Each pixel in
  the images correspond to 0.5 kpc$^{2}$ Clearly, substructure is
  expected from a distribution of M81-like RGB stars. Note that the
  amount of substructure varies from halo to halo, depending on
  accretion history.}
 \label{fig:rgbmaps}
 \end{figure*}

\begin{figure*} \centering
\includegraphics[width=185mm,clip]{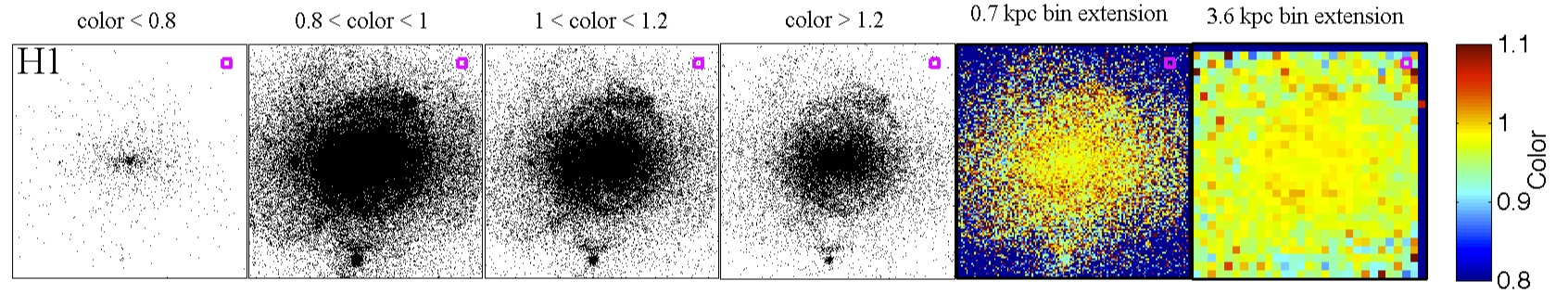}
\includegraphics[width=185mm,clip]{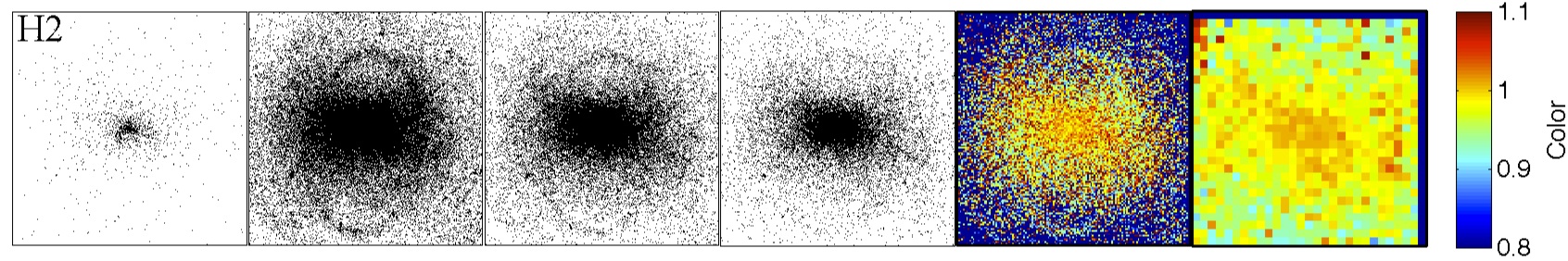}
\includegraphics[width=185mm,clip]{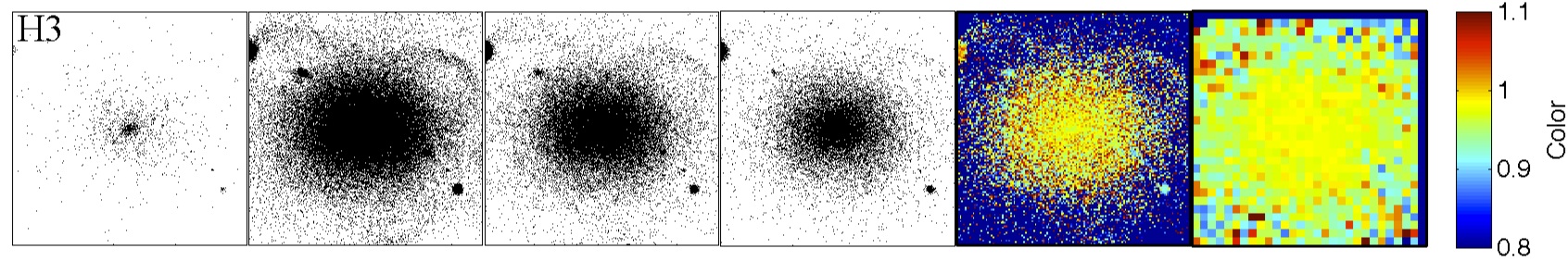} 
\includegraphics[width=185mm,clip]{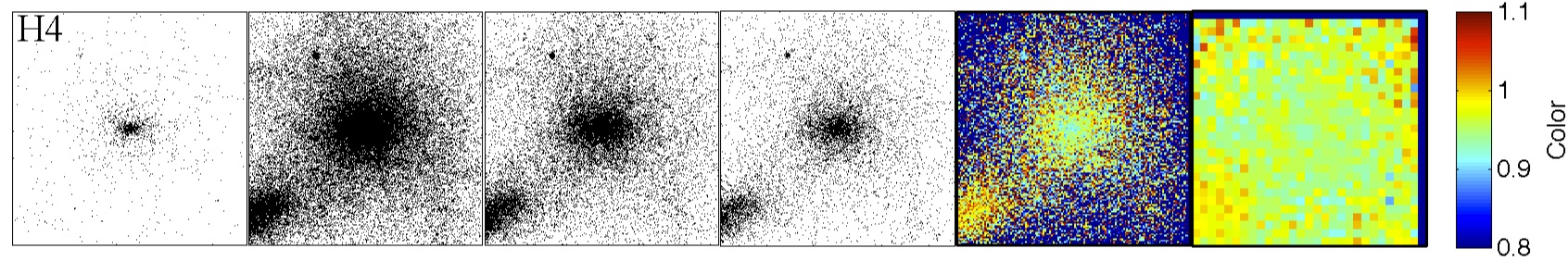}
%%%%%\hspace{0.5cm}
\caption{Color distribution of RGB stars at the distance of M81 for 4
  halo realizations (from top to bottom). From left to right, the
  first four panels show the distribution of RGB stars within
  different color ranges. Some substructures are more dominated by
  either redder or bluer stars than others. The fifth panels show maps
  of the mean colors of all RGB stars with a bin size of 0.5
  kpc$^{2}$. No significant variations in mean colors can be seen
  throughout each map. The right-most panels show the mean color maps
  for a bin size of $3.6\times3.6$ kpc$^2$, resembling the size of the
  ACS/WFC FoV. Note that in this case the substructure has been
  virtually erased. The small degree of variation in the colors is
  challenging to detect with HST given the absolute calibration and
  shot noise uncertainties. The magenta square shown in each of
    the top panels represents the extent of the ACS/WFC FoV at the
    distance of M81.}
 \label{fig:colormaps}
 \end{figure*}

To address this question, we have built RGB maps of the entire halo,
i.e. both smooth component and surviving satellites, at the distance
of M81 considering the limiting magnitude of our HST observations, and
oriented according to M81's inclination. Maps of the density of RGB
stars for 4 different halo models, in a $100\times100$ kpc$^2$ $XZ$
projection box, are shown in Figure~\ref{fig:rgbmaps}. Each pixel in
these images corresponds to an area of 0.5 kpc$^{2}$. From this
figure, it is clear that the models predict a wealth of substructure
in the distribution of RGB stars at M81's distance. Note that the
amount of substructure strongly depends on the accretion history of
each modeled halo. In Figure~\ref{fig:colormaps} we show the mean
color distribution of the RGB stars plotted in
Figure~\ref{fig:rgbmaps}. The first four panels, from left to right
show the distribution of RGB stars on the sky found within different
color ranges. These panels show that some substructures are more
dominated by either redder or bluer stars than others. In the fifth
panels we show maps of the mean colors of all RGB stars, after
considering a bin size of 0.5 kpc$^{2}$, as in
Figure~\ref{fig:rgbmaps}. We notice that there are no significant mean
($F606W-F814W$) color variations throughout each map; each halo seems
to be dominated by one mean color, and departures from this mean are
generally associated with surviving satellite galaxies orbiting the
main halo. The situation becomes even worse when considering a bin
size resembling that of the ACS/WFC FoV, as shown in the right-most
panels. Note that in this case the substructure has been virtually
erased. There is a small degree of variation in the colors, with a
maximal range from $\sim 0.95$ to $\sim1.1$. Such subtle color
differences are challenging to detect with HST given the absolute
calibration and shot noise uncertainties of $\sim$ 0.05 mag in
($F606W-F814W$) color. Note that the inner regions of these maps do
not have median colors as red as the observations at $R<15$. We recall
that the models only simulate halo particles at all radii whereas the
M81 observations have significant contribution from the disk in the
inner $R\sim 15$ kpc. The disk contribution, which is not included in
the BJ05 models, is most likely responsible for the redder median
colors of M81 observed at $R<15$.

\section{Discussion}\label{sec:discussion}

As shown in previous sections, we do not detect a color gradient in
the stellar halo of M81. If we assume that the color profile of the
RGB stars reflects the metallicity profile\footnote{Even though there
  is a mixture of ages in the RGB, the colors of the RGB stars are
  more sensitive to metallicity than to ages.} of the stellar halo,
our results suggest that there is no metallicity gradient in the halo
of M81 within 15 to 50 kpc from the galactic center. We note that
below [Fe/H]$\sim -1$, the expected color variation due to metallicity
changes is small for old populations ($\lesssim$0.2 mag from [Fe/H] $=
-1$ to $\sim -2.5$ dex), thus the data might be also consistent with a
metallicity gradient to lower metallicity. Metallicities obtained
  using spectroscopic lines will, generally, better sample the more
  metal-poor component of an old system. However, we find a median
color change of $0.03\pm0.11$ mag for the RGBs of the halo's fields
within 15 and 50 kpc, which is consistent with no color as well as no
metallicity gradient for old populations. The lack of a metallicity
gradient is in agreement not only with the \citet{BJ05} models
analyzed in this paper, but also with other theoretical studies based
on accretion-only halos which found flat metallicity profiles
\citep[see e.g.,][]{ DeLucia_helmi08, Cooper10, Gomez12}.

On the other hand, \citet{Font11} have analyzed $\sim$ 400
$\textrm{L}_{*}$ disk galaxies from cosmological hydrodynamical
simulations, and found that the average stellar metallicity profile
shows a prominent negative gradient over all radii, with the deepest
decline (a $\sim 0.4$ dex metallicity difference) exhibited over the
range of $20 < R < 40$ kpc. They claim that the slope in the
metallicity profile is induced by 'in situ' star formation, which
typically dominates at $R < 30$ kpc, whereas accretion of stars
dominate at large radii. Our observations of M81 do no support such a
strong gradient. In addition, even if one associates the redder stars
that we observe in M81 (which clearly sample the properties of disk
stars, see their CMDs in Figure~\ref{fig:cmds}), with an 'in situ'
formation, they appear not to extend further than $\sim$15 kpc along
the major axis, and not further than $\sim$10 kpc along the minor
axis. From 15 to 50 kpc, i.e. the region studied in this work,
\citet{Font11} find a 0.7 dex metallicity difference in the spheroidal
component of their simulated galaxies. Since they also find variations
in the mean age with galactocentric distances, we explore whether
their models predict a color gradient of RGB stars when both
metallicity and age profiles are combined. Considering the median
spherically-averaged age and metallicity radial profiles provided by
\citet[see their figures 5 and 10]{Font11}, we find that their results
imply an intrinsic color difference of $\sim$0.3 mag from 15 to 50 kpc
with stellar halo RGB colors ($F606W - F814W$) becoming gradually
bluer with increasing radius. This is not supported by our
observations of M81 from 15 to 50 kpc.

 We can compare our results with observations of halo stars in other
 large galaxies. For the Milky Way, there seems to be a discrepancy
 regarding whether a metallicity gradient exists or not. As discussed
 in Section 1, analysis of SDSS stars by \citet{Carollo07, Carollo10}
 indicated a metallicity drop of $\approx 0.7$ -- 0.8 dex from the
 Solar neighborhood out to $\approx$ 30 -- 40 kpc. A different stellar
 halo sample from recent SDSS/SEGUE observations show a nearly flat
 Galactic halo metallicity distribution from $\sim 20$ kpc out to
 $\sim60$ kpc with an average metallicity of $[\textrm{Fe/H}] \sim
 -1.4$ dex (Ma et al. in prep.). Nevertheless, as mentioned in
 Section~\ref{sec:intro}, these studies may have important biases
 introduced by, e.g., the magnitude or color limit considered, which
 affect the determination of a metallicity--distance relation
 \citep[see e.g.,][]{Schonrich11}. Thus, whether a gradient exists or
 not in the halo of the Milky Way is still unclear.

The metallicity of the stellar halo of M31 has also been extensively
studied. \citet{Kalirai06} and \citet{Koch08} have detected a clear
metallicity gradient with substantial scatter over a large range in
radial distances, from $\approx 10$ kpc with metallicity
$[\textrm{Fe/H}] \sim 0.5$ dex to $\approx 160$ kpc with
$[\textrm{Fe/H}] \sim -1.3$ dex. However, as discussed in
Section~\ref{sec:intro}, photometric studies \citep[see
  e.g.][]{Durrell04, Irwin05, Richardson09} as well as spectroscopic
studies \citep{Chapman06} found no detectable metallicity gradient
from $\approx 10$ to $\approx 60$ kpc. Furthermore, the mean
metallicity values vary from work to work (see Figure 6 of
\citealt{Richardson09}). It is unclear why the results are so diverse;
probably differences arise from analysis of the data using different
methods and techniques as well as from probing different various small
regions of M31's halo, which is known to have a wealth of substructure
and intrinsic metallicity variations. Overall, the inner $\sim 15$ to
50 kpc region of M31's halo seems to exhibit a nearly flat, high
metallicity profile ($[\textrm{Fe/H}] \sim -0.8$ dex), whereas outside
$\sim$ 60 kpc the metallicities are lower ($[\textrm{Fe/H}] \sim -1.3$
dex). It is interesting to note that both \citet{Irwin05} and
\citet{Guhathakurta06} found two different structural components for
the M31's spheroid. Stars at $R \lesssim 30$ kpc exhibit an $R^{1/4}$
power-law (or de Vaucouleurs) surface brightness profile whereas
beyond that radius and out to $\sim 160$ kpc, the surface brightness
profile flattens considerably and can be better fitted with a
$R^{-2.3}$ power-law. \citet{Kalirai06} argue that the metal-rich and
metal-poor components are respectively associated with the two
different structural components.

Stellar halos of large early type (E/S0) galaxies have also been
studied. The halos of NGC 5128 and NGC 3377 show no metallicity
gradients out to large galactocentric distances of $\sim 40$ kpc
($\sim 7 R_{e}$) for NGC 5128 \citep{Rejkuba05}, and $\sim 18$ kpc
($\sim 4R_e$) for NGC 3377 \citep{Harris07a} and both their mean
metallicities are $[\textrm{Fe/H}]\sim -0.6$ dex. They contain
virtually no stars more metal-poor than $[\textrm{Fe/H}]\sim -1.5$
dex. \citet{Harris07b}, on the other hand, studied the giant
elliptical galaxy NGC 3379 and indicated the presence of a mild
metallicity gradient, with low metallicity stars ($[\textrm{Fe/H}]
\lesssim -0.7$ dex) dominating in the outermost parts of their field,
which reaches a projected distance of $\sim 12R_e$. In units of
effective radius, this field is farther out than those analyzed for
NGC 5128 and NGC 3377. \citet{Harris07b} thus suggest that large early
type galaxies will have a diffuse, low metallicity halo component
detectable at radius larger than $\sim 10 R_e$ from the galactic
center.

 The metallicity profiles observed for all the galaxies discussed, and
 the color profile detected in this work for M81, can be reproduced by
 simulated halos built entirely from accreted
 satellites. \citet{Cooper10} find a diversity of metallicity gradient
 behaviors in their models, ranging from the lack of a detectable
 gradient to some systems with gradients or breaks or jumps in their
 metallicity profiles. Overall, there is little or no metallicity
 gradient when many satellites contribute comparably to the final
 halo, while metallicity profiles show steeper gradients or present
 sharp variations when only one or two massive satellites contribute
 significantly to the halo.

\emph{In short, the stellar halos of large galaxies display a great
  diversity of metallicity profiles, which reflects the stochasticity
  of halo mass assembly formation history.}
 
\section{Summary and Conclusions}\label{sec:summary}

We have analyzed 19 HST ACS fields in the outer disk and stellar halo
of M81 from the GHOSTS survey, where the term M81's halo in this work
denotes the faint, extended $R^{-2}$ structural component detected by
B09 and confirmed using the GHOSTS fields, which start to dominate the
light at $R > 15.5$ kpc, i.e. $\approx 6$ scale lengths (B09). These
fields probe the stellar halo of M81 out to a projected radial
distance as large as $\sim$ 50 kpc from the galactic center. We have
derived the color function (CF) distributions of each of these
fields. Fields closer than $R \sim 15$ kpc display redder median
colors ($F606W - F814W \sim 1.3$) and typically wider CFs (width $
\sim 0.8$ mag). A visual inspection of their CMDs, as well as their
color profile shape as a function of major axis (see top panel of
Figure~\ref{fig:mod-data1}), suggests that these fields are dominated
by metal-rich disk stars. Fields located at $R > 15.5$ have nearly
constant median color, consistent with being dominated by more-metal
poor halo stars. {\it We do not detect any color gradient within the
  halo of M81 from 15 kpc out to 50 kpc}. There is, nevertheless, a
small degree of scatter in the colors from field to field, but it is
unclear whether this scatter is physical or is due to instrumental
uncertainties.

The halo of M81 is characterized by a color distribution of width
$\sim 0.4$ mag and an approximately constant median ($F606W-F814W$)
$\sim 1$ mag, fluctuating by less than $\pm 0.06$ mag over a range of
$\sim$35 kpc. If we fit for a color gradient, we obtain a slope of
$-0.0009\pm0.0031$ mag/kpc, which places a limit of $0.031\pm0.11$ mag
in the difference between the median color of RGB M81 halo stars at
$\sim$15 and at 50 kpc. This color gradient corresponds to a
difference of $0.08\pm0.35$ dex in [Fe/H] over that radial range, if
we assume a constant age of 10 Gyr.

 We directly compared these results, assuming that what we observe is
 indeed the stellar halo of M81, with predictions by simulations of
 the formation of stellar halos, using the cosmologically motivated
 models provided by \citet{BJ05}. From their predicted stellar
 populations which cover a wide range of ages and metallicities, we
 built synthetic CMDs of ACS-like fields, rotated by $60^{\circ}$ to
 resemble M81's inclination, at different locations from 5 to 50
 kpc. We simulated the observational effects to make a fair comparison
 with the observed data. The model-CMDs display a narrow RGB, despite
 the mixture of populations predicted by the models.

 After analyzing the synthetic stars in the same way as the data, we
 find that there is no color gradient in the models. The average color
 is ($F606W-F814W$) $\sim 1.1$ mag with a small degree of scatter
 within one halo model, as well as from model to model. This lack of a
 gradient is in very good agreement with the observations of M81 for
 fields at $R > 15$ kpc where the contamination from metal rich disk
 stars becomes negligible. The widths of the CFs constructed from the
 models, which correspond to a spread in metallicity of the order of
 $\sim 0.6$ dex are also in good agreement with the observations for
 these fields. This suggests that the observed M81 halo fields contain
 a similar spread in metallicity.

Since this work is based on pencil-beam observations, we investigated
our sensitivity to detecting substructure with color distributions.
For this purpose, we have built RGB star maps of different model halos
at the distance of M81, considering the limiting magnitude imposed by
HST. We do find no significant mean color variations throughout these
maps. Most of each halo appears to be dominated by one mean color and
variations are generally related to stellar populations of surviving
satellite galaxies orbiting the main halo. Much of this subtlety in
color structure is driven by the age-metallicity anticorrelation of
stars in the model halos, combined with the modest sensitivity of RGB
($F606W-F814W$) colors to variations in age and metallicity of the
magnitude predicted by the models. This illustrates that measuring
stellar population variations using RGB ($F606W-F814W$) colors in
stellar halos will require stringent calibration and shot noise
uncertainties. We note, nevertheless, that the sensitivity of the RGB
colors to variations in metallicity could be improved if a wider color
baseline is used.

Finally, we note that our non detection of any significant color (and
likely metallicity) gradient in M81's halo agrees well with many of
the studies of M31's halo {\it within} the same galactocentric
distances, i.e., from $\sim 15$ to $\sim 50$ kpc, as well as with the
Ma et al. (in prep.) study on the halo of our own Milky Way. The
median metallicity that we find for M81's halo, $[\textrm{Fe/H}]\sim
-1.2$ dex, however, is more metal poor than that of the M31 halo
($[\textrm{Fe/H}]\sim -0.8$), but more metal rich than that of the
Milky Way stellar halo ($[\textrm{Fe/H}]\sim -1.6$), as measured
between $\sim 15$ to $\sim 50$ kpc from the center of each galaxy. The
observed differences are possibly due to the different assembly
history that these galaxies may have had.  Given that the Milky Way,
M31 and M81 have similar luminosities \citep{Karachentsev05,
  Mouhcine05b}, their different halo's metallicities seem to indicate
a disagreement with the halo's metallicity -- parent galaxy luminosity
correlation reported by \citet{Mouhcine05b}. These results, on the
other hand, are in agreement with the simulations by \citet{Renda05},
who predicted that at any given luminosity the metallicities of their
simulated stellar halos (89 in total) span a range of $\approx 1$ dex
(see their Figure 1). They suggest that the diversity in halo
metallicities arise from the differences in the galactic mass assembly
histories.

The stochasticity of galaxy formation in a cosmological context
results in a great diversity between the formation histories of the
stellar halos.  Large, nearby spiral galaxies appear to exhibit a wide
variety of halo metallicities~\citep[see e.g.][]{Mouhcine05c,
  Rejkuba09}.  The GHOSTS data set provides important tests for
current models of galaxy formation and evolution, since it enlarges
the number of observed spiral galaxy's halos, required to obtain a
more statistically significant sample for comparison with the
different models.  As shown in Figure~\ref{fig:mod-data1}, small color
variations (due to age and metallicity variations) among different
predicted halos are expected and different halo metallicities are also
expected due to differences in the galactic mass assembly histories
\citep[see e.g.][]{Renda05, Robertson05, Cooper10, Gomez12}. In a
follow-up paper, we will analyze the rest of the galaxies observed
within the GHOSTS survey and we will quantify the scatter in the color
profiles from observed halos as well as investigate whether there are
correlations between the color or metallicity of the halo stars and
the main halo properties, such as mass or morphological type.

\acknowledgments We are grateful to Tom Brown, Anil Seth, and Henry
Ferguson for their valuable comments and suggestions.  We wish to
thank the anonymous referee for the careful reading of our manuscript
and comments that helped improve this paper. This work was supported
by HST grant GO-11613, provided by NASA through a grant from the Space
Telescope Science Institute, which is operated by the Association of
Universities for Research in Astronomy, Inc., under NASA contract
NAS5-26555. This work has made use of the IAC-STAR synthetic CMD
computation code. IAC-STAR is supported and maintained by the computer
division of the Instituto de Astrof\'isica de Canarias. We acknowledge
the usage of the HyperLeda database.

\emph{Facility:} \facility{HST (ACS)}

\bibliographystyle{apj}
\bibliography{ref}

\end{document}